\documentclass[journal,twoside,twocolumn,10pt]{IEEEtran}
\usepackage{cite}

\ifCLASSINFOpdf
\else
  \usepackage[dvips]{graphicx}
\fi
\usepackage{color}

\usepackage[cmex10]{amsmath}
\usepackage{amssymb}
\usepackage{multirow}
\usepackage{amsthm}
\allowdisplaybreaks[4]

\usepackage[tight,footnotesize]{subfigure}
\usepackage{url}

\newcommand{\ignore}[1]{}

\newtheorem{lemma}{Lemma}
\newtheorem{theorem}{Theorem}

\newtheorem{corol}{Corollary}

\begin{document}
%
\title{Coverage and Throughput Analysis with a Non-Uniform Small Cell Deployment}
%
%
%
\author{He~Wang,~\IEEEmembership{Member,~IEEE,}
        Xiangyun~Zhou,~\IEEEmembership{Member,~IEEE,}
        Mark~C.~Reed,~\IEEEmembership{Senior Member,~IEEE}
\thanks{This work was supported by the Australian Research Council's Discovery Projects funding scheme (Project No. DP110102548 and DP130101760) and National ICT Australia. A part of this paper has been presented at IEEE International Conference on Communications (ICC'2013) in Budapest, Hungary \cite{WanZho13ICC}.}
\thanks{H. Wang was with the Australian National University and National ICT Australia at the time of this writing. He is now with the School of Engineering and Information Technology, University of New South Wales (UNSW) Canberra, ACT 2600, Australia (e-mail: jackson.wang@unsw.edu.au).}
\thanks{X. Zhou is with the Research School of Engineering, the Australian National University, ACT 0200, Australia (e-mail: xiangyun.zhou@anu.edu.au).}
\thanks{M. C. Reed is with the School of Engineering and Information Technology, UNSW Canberra, ACT 2600, Australia, and also with the Research School of Engineering, the Australian National University, ACT 0200, Australia (e-mail: mark.reed@unsw.edu.au).}
}

\ifCLASSOPTIONdraftclsnofoot
\else
\fi
%



\maketitle

\begin{abstract}
  Small cell network (SCN) offers, for the first time, a low-cost and scalable mechanism to meet the forecast data-traffic demand. In this paper, we propose a non-uniform SCN deployment scheme. The small cell base stations (BSs) in this scheme will not be utilized in the region within a prescribed distance away from any macrocell BSs, defined as the inner region. Based upon the analytical framework provided in this work, the downlink coverage and single user throughput are precisely characterized. Provided that the inner region size is appropriately chosen, we find that the proposed non-uniform SCN deployment scheme can maintain the same level of cellular coverage performance even with $50$\% less small cell BSs used than the uniform SCN deployment, which is commonly considered in the literature. Furthermore, both the coverage and the single user throughput performance will significantly benefit from the proposed scheme, if its average small cell density is kept identical to the uniform SCN deployment. This work demonstrates the benefits obtained from a simple non-uniform SCN deployment, thus highlighting the importance of deploying small cells selectively.
\end{abstract}

\ifCLASSOPTIONdraftclsnofoot
\else
  \begin{IEEEkeywords}
    Small cell networks, non-uniform deployment, coverage, throughput, stochastic geometry.
  \end{IEEEkeywords}
\fi

%
\IEEEpeerreviewmaketitle

\section{Introduction}\label{sec:Introduction}
%
%
%
%

In recent years, the cellular communications industry has experienced an unprecedented growth in the numbers of subscribers and data traffic. This significant trend challenges cellular service providers' traditional macro-only network: A much more advanced and flexible network topology is desired. To meet this demand, the concept of heterogeneous network is proposed to most efficiently use the dimensions of space and frequency. Its network topology is composed of a diverse set of wireless technologies, traditional macrocells and low-power small cells \cite{And13MCOM}. By off-loading wireless traffic from macro to small cells and decreasing the distance from users to base stations (BSs), small cell network (SCN) bring a multitude of benefits, including improved user experiences and more efficient spatial reuse of spectrum~\cite{AndCla12JSAC}. \ignore{In this work, we focus on the open-access femtocells, which are operated by the cellular service providers and offer femtocell access to all the users in the networks \cite{Sau09Book, XiaCha10TWC}.}

The cellular coverage performance of a SCN strongly depends on the locations of small cell BSs. With a constant pre-configurable transmit power, which is a mode commonly implemented in current solutions \cite{ChaKou09TWC, WeiGro10MCOM}, the small cell coverage range is significantly reduced when it is close to a macrocell BS site \cite{JoSan12TWC}, resulting in poor off-loading effects. More interestingly, when the small cell BSs are uniformly deployed at random, increasing the density of small cell BSs does not give any noticeable improvement in the coverage probability \cite{JoSan12TWC, DhiGan12JSAC, WanRee12AusCTW}. The main cause of this phenomenon is the increased network interference from having more small cell BSs in satisfactory macrocell areas. Hence, one interesting question raised from the above-mentioned discussion is whether or not we can improve both coverage and throughput performances by not utilizing the small cell BSs at undesirable locations, in other words, deploying small cells non-uniformly.

\begin{figure}[t!]
  \centering
  \includegraphics[height=0.26\textheight, bb=140 262 443 563, clip = true]{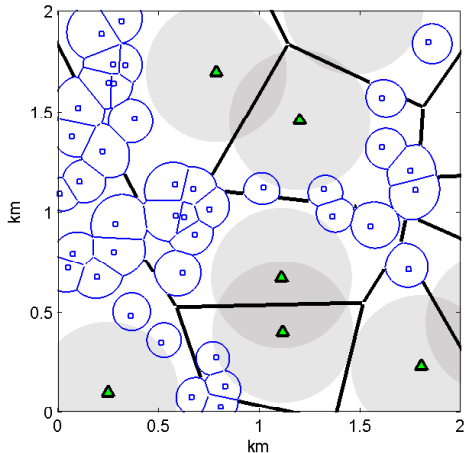}
  \caption{Illustration of cell association regions for the proposed non-uniform SCN deployment scheme. Macrocell BSs (denoted by triangles) are randomly scattered in the whole plane. Small cell BSs (also randomly scattered and denoted by squares) are only utilized in the outer region, and not used in the inner region (denoted by shadow areas).}
  \label{fig:CoverageOriented_Deployment_Demo} 
\end{figure}

In our analysis, the union of locations within a prescribed distance from any macrocell BSs is defined as the \emph{inner region}, shown as the shadow areas in Fig.~\ref{fig:CoverageOriented_Deployment_Demo}, whereas the union of locations outside the inner region is defined as the \emph{outer region}. Here, we consider an intuitive and interesting idea: We simply avoid using small cell BSs within the inner region, illustrated in Fig.~\ref{fig:CoverageOriented_Deployment_Demo}. It is expected that the small cell locations are known to the cellular operator which uses provisioning processes to avoid small cells deployment in certain regions. This actually occurs today with the operators only assigning femtocell access points to particular sites based on system constraints, where our non-uniform SCN deployment scheme can be regarded as one way to achieve that.

\subsection{Approach and Contributions}
In this work, we aim to show the impact of employing the proposed non-uniform SCN deployment on the downlink coverage and throughput performance of the two-tier cellular network. Specifically, our goal is to derive the coverage probability, or equivalently the distribution of signal-to-noise ratio (SINR), based on which the throughput achievable at a randomly chosen user can further be derived.

Fortunately, modeling BSs to be randomly placed points in a plane and utilizing stochastic geometry \cite{StoKen95Book, BacBla09Book} to study cellular networks has been used extensively as an analytical tool with improved tractability \cite{Bro00JSAC, YanPet03TSP, Hae08TIT}. Recent works \cite{AndBac11JCOM, NovGan11TWC, JoSan12TWC, DhiGan12JSAC, WanQue12JSAC, CheNgu12VTC, YuKim11arXiv, SinDhi12arXiv} have shown: Compared with the practical network deployment, modeling the cellular network with BS locations drawn from a homogeneous Poisson Point Process (PPP) is as accurate as the traditional grid models. Moreover, the stochastic geometry model can provide the randomness introduced by the small cell deployment, and also more tractable analytical results on both the coverage and throughput performances. Based on these reasons, this stochastic geometry tool is adopted to model the locations of BSs in this work.

The main contributions of this paper are as follows:
\begin{enumerate}
  \item We propose the above-mentioned non-uniform SCN deployment scheme. In this scheme, small cell BSs are not utilized in the region within a certain distance away from any macrocell BSs. Through this scheme, we can guarantee that most of the active small cell BSs are located in the relatively poor macrocell coverage areas.
  \item By employing the PPP-based stochastic-geometry BS model in the analysis, we provide the probabilistic characterization of the downlink coverage and single user throughput achievable by a randomly located mobile user in this new scheme. To our knowledge, it is the first study to derive analytical results on a non-uniform SCN deployment.
  \item Compared with the uniform SCN deployment, we demonstrate that the same cellular coverage performance can be maintained if the size of the inner region for the proposed scheme is appropriately chosen. Our numerical result demonstrates that the number of utilized small cell BSs can be reduced by more than $50$\%, without compromising the coverage performance.
  \item By maintaining the average small cell density in the proposed scheme, we show that both coverage and single user throughput can be significantly improved over the uniform SCN deployment, namely $72$\% improvement for the throughput achievable by the worst $10$\% users. This finding emphasizes the importance of deploying the small cells selectively by taking their relative locations with macrocell BSs into account.
\end{enumerate}

It should be noticed that Haenggi introduced a non-uniform multi-tier deployment model that also incorporates dependencies between different cellular tiers \cite{Hae13arXiv}. The superposed tiers in Haenggi's model are deployed on the edges and at the vertices of the Voronoi cells formed by the macrocell BSs, that is, the poorest macrocell coverage locations. In contrast, our scheme regards the previously defined outer region as the poor macrocell coverage area, in which small cell BSs can be deployed. Additionally, our study provides not only the new model, but also the analysis on both coverage and throughput performance.

The remainder of the paper is organized as follows: In Section~\ref{sec:SysModel}, the system model and the analysis on the uniform SCN deployment used in this work are introduced. Section~\ref{sec:CovDeployment} provides the tractable result for the downlink coverage and the achievable single user throughput of the proposed non-uniform SCN deployment scheme. Section~\ref{sec:NumResults} presents numerical results and we conclude the paper in Section~\ref{sec:Conclusion}.


\section{System Model and Uniform SCN Deployment}\label{sec:SysModel}
\subsection{Two-Tier Cellular Network Model}\label{subsec:NetworkModel}

We consider the downlink environment of a heterogeneous cellular network (HetNet), which employs an orthogonal multiple access technique, like the orthogonal frequency-division multiple access (OFDMA) in Long Term Evolution (LTE). In this analysis, the HetNet consists of two tiers, the macro and small cell tiers (otherwise called as the first and second tiers), which are spatially distributed as two-dimensional processes $\Phi_{1}$ and $\Phi_{2}$, with different transmit powers $P_{tx,1}$ and $P_{tx,2}$ respectively. The same transmit power holds across each tier. The macrocell tier process $\Phi_{1}$ is modeled as a homogeneous PPP with the density $\lambda_1$. Furthermore, the collection of mobile users, located according to an independent homogeneous PPP $\Phi_{MS}$ of the density $\lambda_{MS}$, is assumed in this work. We consider the process $\Phi_{MS} \cup \{0\}$ obtained by adding a user at the origin of the coordinate system, which is the typical user under consideration. This is allowed by Slivnyak's Theorem \cite{StoKen95Book}, which states that the properties observed by a typical point of the PPP $\Phi_{MS}$, are identical to those observed by the origin in the process $\Phi_{MS} \cup \{0\}$.

Here we use the standard power loss propagation model with path loss exponent $\alpha > 2$ and path loss constant $L_0$ at the reference distance $r_0 = 1$~m. We assume that the typical mobile user experiences Rayleigh fading from the serving and interfering BSs. The impact of fading on the signal power follows the exponential distribution with the unitary mean value\ignore{, i.e. $h \sim \exp(1)$}. This Rayleigh fading model has been proven to be representative for all kinds of shadowing/fading, because PPP-based networks with arbitrary distributions of shadowing/fading have identical distributed signals perceived at a given location as long as the shadowing/fading values have the same value of the $2/\alpha$-moment \cite{BlaKar13Infocom}. The noise power is assumed to be additive and constant with value $\sigma^2$.

\subsection{Cell Association, Traffic and Resource Allocation Model}\label{subsec:CellAssociation}
As we described earlier, all macrocell and small cell BSs are open access, and there is no limitation on the number of users served by each BS. Moreover, we assume that mobile users are connected to the BS providing maximum long-term received power, which can be regarded as a widely used special case of the general cell association model \cite{AndBac11JCOM, JoSan12TWC}. Specifically, the selected tier index $\kappa$ is
  \begin{equation}\label{eqn:omega}
    \kappa = \arg \max_{i \in \{1,2\}} \big[ {P_{tx,i} L_0 (R_i)^{-\alpha}} \big],
  \end{equation}
in which $R_i$ is the minimum distance from the $i$-th tier BSs to the typical mobile user at the origin, that is, $R_i = \min_{x \in \Phi_{i}} \|x\|$.
For simplicity, we use $P_i$ to replace the product of $P_{tx,i}$ and $L_0$, i.e., $P_i = P_{tx,i} L_0, i \in \{1,2\}$, from now on.

We utilize the full buffer traffic model for all the active users \cite{SinDhi12arXiv, NiuLee13TWC}, in which the buffers of the users' data flows always have unlimited amount of data to transmit. Best effort traffic is assumed, and there is no quality and latency requirements for the traffic model. The available resources at each BS are assumed to be allocated evenly among all its mobile users to simulate the Round-Robin scheduling with the most fairness.

\subsection{Received SINR for Data Channels}\label{subsec:SINR}
In this work, we specifically focus on the data channels, an important additional feature which is different from other previous works in the field. This better characterizes practical scenarios: When a BS has no user to serve, all the frequency-time resource blocks for data channels will be left blank, while the other BSs with at least one user associated will occupy all the resource blocks for data transmissions \cite{DahPar08HSPALTEbook}. Therefore, there is the likelihood that there is no user for a BS to serve and thus no data-channel interference is generated from this BS in our analysis.
We use $\{\Phi'_i\}_{i \in \{1,2\}}$ to denote the process constructed by the remaining $i$-th tier BSs (i.e., the loaded BSs in the $i$-th tier) excluding the ones not associated with users (i.e., unloaded BSs).

By employing the cell association model described in Subsection~\ref{subsec:CellAssociation} to choose the serving cell, the downlink received SINR at the typical user can be expressed as
  \begin{equation}\label{eqn:SINR}
    \mathrm{SINR} = \frac{P_\kappa h (R_{\kappa})^{-\alpha}}{I + \sigma^2},
  \end{equation}
where $I = \sum_{i \in \{1,2\}} I_i = \sum_{i \in \{1,2\}} \sum_{x \in \Phi'_i \setminus \{x_o\}} P_i  h_{x} \|x\|^{-\alpha}$ is the cumulative interference from all the loaded BSs (except the BS at $x_o$ from the $\kappa$-th tier serving for the mobile user at $o$), and $I_i$ is defined as the interference component from the $i$-th tier. Here, $h$ is employed to denote the channel fading gain from the serving BS, and $h_x$ is the value for the interfering BS at the location of $x$. Based upon the Rayleigh fading assumption, all these channel fading gains, $h$ and $\{h_x: x \in \Phi'_i \setminus \{x_o\}\}$ follow the exponential distribution with the unitary mean value. In our study, no intracell interference is incorporated since we assume that the orthogonal multiple access is employed among intracell users.


\subsection{Uniform SCN Deployment}\label{subsec:UniformDeployment}
Before proposing the non-uniform SCN deployment scheme, we firstly focus on the uniform SCN deployment, where the small cell BSs in $\Phi_2$ are located as a homogeneous PPP with the density $\lambda_2$ in the whole plane. Following the analysis in \cite{JoSan12TWC, SinDhi12arXiv}, the results for this uniform SCN deployment are briefly presented here for the purpose of comparison.

\subsubsection{Mobile User Resource Sharing}
Firstly, the per-tier association probabilities for the uniform SCN deployment, that is, the probabilities for the typical user to associate with macrocell and small cell tiers, denoted by $\mathcal{Q}_{1,u}$ and $\mathcal{Q}_{2,u}$, were derived in \cite{JoSan12TWC}, i.e.,
  \begin{align}\label{eqn:prob_tier1_tier2_unifemto}
    \mathcal{Q}_{1,u} = \frac{\lambda_1}{\lambda_1 + \lambda_2 \big(\frac{P_2}{P_1}\big)^{{2}/{\alpha}}} \ \text{    and    } \  \mathcal{Q}_{2,u} = \frac{\lambda_2}{\lambda_1 \big(\frac{P_1}{P_2}\big)^{{2}/{\alpha}} + \lambda_2}.
  \end{align}

As proved in \cite{SinDhi12arXiv}, the area of the $i$-th tier cells, denoted by $\mathcal{C}_{i,u}$, can be well approximated by the Voronoi cell area formed by a homogeneous PPP with the density value $\lambda_i/\mathcal{Q}_{i,u}$, i.e., $\mathcal{C}_{i,u} \approx \mathcal{C}_0 ({\lambda_i}/{\mathcal{Q}_{i,u}})$, in which $C_0(y)$ is the area of a typical Voronoi cell of a homogeneous PPP with the density $y$. For the distribution of Voronoi cell area formed by a homogeneous PPP, there is no known closed form expression for its distribution \cite{OkaBoo92Book}; however, some precise estimates can be conducted \cite{HinMil80Math,WeaKer86Math}: the approximated probability density function (PDF) of $\mathcal{C}_{i,u}$ can be expressed as
  \begin{align}\label{eqn:CellArea_RC_pdf}
    f_{\mathcal{C}_{i,u}}(x) \approx (b \lambda_{i,eq,u})^q x^{q-1} \exp(-b \lambda_{i,eq,u} x)/\Gamma(q),
  \end{align}
where $q = 3.61$, $b = 3.61$, $\lambda_{1,eq,u} = \lambda_1 + \lambda_2 ({P_2}/{P_1})^{{2}/{\alpha}}$, $\lambda_{2,eq,u} = \lambda_1 ({P_1}/{P_2})^{{2}/{\alpha}} + \lambda_2$ and $\Gamma(x) = \int_0^{\infty} t^{x-1} e^{-t} \mathrm{d} t$ is the standard gamma function.

Based on this approximation, the probability mass function (PMF) of the number of users in a randomly chosen $i$-th tier cell can be derived as
  \begin{multline}\label{eqn:PMF_MS_num_unifemto_RC}
    \mathbb{P}[N_{i,c,u} = n]
    \approx \frac{b^q}{n!} \cdot \frac{\Gamma(n+q)}{\Gamma(q)} \cdot \frac{(\lambda_{MS})^n (\lambda_{i,eq,u})^q}{(\lambda_{MS} + b \lambda_{i,eq,u})^{n+q}}, \\ \ \text{ for } n \in \mathbb{Z}^0 \text{ and } i \in \{1,2\}.
  \end{multline}

Provided that the typical user is enclosed in the cell, the PDF of the $i$-th tier cell area $\mathcal{C}_{i,u}$ was derived in \cite{SinDhi12arXiv,YuKim11arXiv}, that is, $f_{\mathcal{C}_{i,u} \mid o \in \mathcal{C}_{i,u}}(x) = {x f_{\mathcal{C}_{i,u}}(x)}/{\mathbb{E}[\mathcal{C}_{i,u}] }$
which helps to obtain the distribution of the number of in-cell users sharing the resource with the typical user
  \begin{multline}\label{eqn:PMF_MS_num_unifemto}
    \mathbb{P}[N_{i,u} = n] \approx \frac{b^q}{n!} \cdot \frac{\Gamma(n+q+1)}{\Gamma(q)} \cdot \frac{(\frac{\lambda_{MS}}{\lambda_{i,eq,u}})^n}{(b + \frac{\lambda_{MS}}{\lambda_{i,eq,u}})^{(n+q+1)} },
    \\ \text{\ \ \ for } n \in \mathbb{Z}^0, i \in \{1,2\}.
  \end{multline}

As the cell coverage regions are mutually disjoint and PPP $\Phi_{MS}$ has the property of complete independence \cite{StoKen95Book}, the numbers of users in different cells are independent. For a randomly chosen $i$-th tier cell, its probability to be an unloaded cell is $\mathbb{P}[N_{i,c,u} = 0]$. Hence, the process $\Phi'_i$, (i.e., the $i$-th tier loaded BSs excluding the BSs without users associated,) can be approximated by a homogeneous PPP with the density $\lambda'_{i,u} = \lambda_i \cdot (1 - \mathbb{P}[N_{i,c,u} = 0])$.

\subsubsection{Coverage Probability}
We use $\mathrm{SINR}_{i,u}$ to denote the received SINR at the typical user served by the $i$-th tier for this uniform SCN deployment. Then the coverage probability at the typical user is $p_{c} (T) = \mathbb{P}[\mathrm{SINR}_{i,u} > T ]$ for the $i$-th tier, i.e., the probability of a target SINR $T$ (or SINR threshold) achievable at the typical user. This coverage probability is also exactly the complementary cumulative distribution function (CCDF) of the received SINR.

If the typical user is served by the $i$-th tier, its coverage probability has been derived in~\cite{JoSan12TWC}:
  \begin{align}\label{eqn:probcov_unifemto}
    p_{c, i, u} (T) & = \mathbb{P}[\mathrm{SINR}_{i,u} > T ] \nonumber \\
    &= 2\pi \lambda_{i,eq,u} \int_{x>0} x \exp (-\frac{T x^{\alpha} \sigma^2}{P_i} ) \nonumber \\
    & \ \ \ \ \ \ \ \cdot \exp \bigg[- \pi x^2 \Big[ \lambda_{i,eq,u} +\rho(T,\alpha) \lambda'_{i,eq,u} \Big] \bigg] \mathrm{d}x, \nonumber \\
    & \ \ \ \ \ \ \ \ \ \ \ \ \ \ \ \ \ \ \ \ \ \ \ \ \ \ \ \ \ \ \ \ \ \ \text{   for } i \in \{1,2\},
  \end{align}
where the function $\rho(x,\alpha)$ is defined as $\rho(x,\alpha) = x^{2/\alpha} \int_{x^{-2/\alpha}}^{\infty} \big({1}/{(1+u^{\alpha/2})}\big) \mathrm{d}u$, and $\{\lambda'_{i,eq,u}\}_{i \in \{1,2\}}$ can be defined as $\lambda'_{1,eq,u} = \lambda'_{1,u} + \lambda'_{2,u} ({P_2}/{P_1})^{{2}/{\alpha}}$ and $\lambda'_{2,eq,u} = \lambda'_{1,u} ({P_1}/{P_2})^{{2}/{\alpha}} + \lambda'_{2,u}$. 

\subsubsection{Single User Throughput}
Following the analysis in \cite{SinDhi12arXiv}, the throughput achievable at the typical user served by the $i$-th tier, denoted by $\mathcal{R}_{i,u}$, can be derived as
  \begin{align}\label{eqn:Thput_unifemto}
    \mathbb{P}[\mathcal{R}_{i,u} > \rho] & = \mathbb{P}\Big[\frac{W}{N_{i,u} + 1} \log_2(1+\mathrm{SINR}_{i,u}) > \rho \Big] \nonumber \\
    & \approx \sum_{n=0}^{\infty} \mathbb{P}[N_{i,u} = n] \cdot p_{c, i, u} \big(2^{(n + 1)\rho /W} -1 \big),  \nonumber \\
    & \ \ \ \ \ \ \ \ \ \ \ \ \ \ \ \ \ \ \ \ \ \ \ \ \ \ \ \ \ \ \text{   for } i \in \{1,2\},
  \end{align}
where the BS's bandwidth $W$ are evenly allocated among all its associated users, namely, the typical user and the other $N_{i,u}$ in-cell users, as stated in the resource allocation model mentioned earlier. It should be noted that the approximation is achieved by assuming the independence between the distribution of $\mathrm{SINR}_{i,u}$ and $N_{i,u}$, and this assumption was proved to be accurate \cite{SinDhi12arXiv}.

\section{Non-Uniform SCN Deployment}\label{sec:CovDeployment}
In this section, we analyze the proposed non-uniform SCN deployment scheme, which aims to make all the small cell BSs utilized in the areas with unsatisfactory macrocell coverage. Based on our analysis using the tool of the stochastic geometry cellular network model, the main results are the probabilistic characterizations of the downlink coverage and the achievable single user throughput presented in Subsections~\ref{subsec:covprob} and \ref{subsec:single_user_throughput}, respectively. Before introducing the main results, Subsection~\ref{subsec:inner_outer_region} provides the definition of the inner and outer regions, and Subsection~\ref{subsec:loaded_BS} provides the analysis on the process of loaded BSs.

\subsection{Inner and Outer Regions}\label{subsec:inner_outer_region}
By implementing the non-uniform SCN deployment scheme, the small cell BSs will be only utilized in the inner region. Specifically, the inner region $A_{inner}$ are defined as the union of locations in which the distance to the nearest macrocell BS site is no larger than $D$, and the outer region $A_{outer}$ are defined as the union of locations whose distances to any macrocell BSs are larger than $D$, that is,
  \begin{equation}\label{eqn:A}
    A_{inner} = \bigcup_{x \in \Phi_1} B(x,D) \text{ and } A_{outer} = \mathbb{R}^2 \setminus A_{inner},
  \end{equation}
where $D$ is called the radius of inner region in this paper. Following PPP's void probability~\cite{StoKen95Book}, the typical user's probabilities to be  located in the inner region $A_{inner}$ and the outer region $A_{outer}$ are
  \begin{align}\label{eqn:Prob_A_inner}
    \mathbb{P}[o \in A_{inner}] = 1 - \exp(-\pi \lambda_1 D^2),
  \end{align}
and
  \begin{align}\label{eqn:Prob_A_outer}
    \mathbb{P}[o \in A_{outer}] = \exp(-\pi \lambda_1 D^2).
  \end{align}

Therefore, the small cell BSs are distributed according to the homogeneous PPP with the density $\lambda_2$ in the outer region only, and don't be utilized in the inner region, thus making the small-cell-tier process $\Phi_2$ become a \emph{Poisson hole process}\cite{LeeHae12TWC}, and the average density for the small cell BSs over the whole plane become $\mathbb{E}[\lambda_{2}(x)] = \lambda_2 \cdot \mathbb{P}[o \in A_{outer}]$.

\subsection{The Density of Loaded BSs} \label{subsec:loaded_BS}
As stated in Section~\ref{sec:SysModel}, we assume that the interference signal at the data channel is only generated from the loaded BSs (i.e., the BSs having at least one user associated). We use the process $\Phi'_i$ to denote the $i$-th tier loaded BSs for the non-uniform SCN deployment scheme, excluding the unloaded ones without users associated. Like the method used in Subsection~\ref{subsec:UniformDeployment}, we assume that the process $\Phi'_i$ can be approximated by a homogeneous PPP with the density $\lambda'_{i} = \lambda_i \cdot (1 - \mathbb{P}[N_{i,c} = 0])$, where $N_{i,c}$ is the number of users in a randomly chosen $i$-th tier cell. We are interested in the density of $\Phi'_i$ over its deployment region, i.e., $\lambda'_i$, which will be used to estimate the interference process later on.

\begin{lemma} \label{lemma:active_density}
The density of $\Phi'_i$ over the corresponding macro- and small-cell-tier deployment regions can respectively be approximated by
  \begin{align}\label{eqn:active_density}
    & \lambda'_1 \approx \lambda_1 \bigg[1 - \Big(\frac{ \lambda_1 b}{\lambda_{MS} \mathcal{Q}_{1} + b \lambda_1} \Big)^q \bigg] \nonumber \\
    \text{  and  } & \lambda'_2 \approx \lambda_2 \bigg[1 - \Big(\frac{ \lambda_2 b}{\lambda_{MS} \mathcal{Q}_{2,outer} + b \lambda_2} \Big)^q \bigg],
  \end{align}
where $\mathcal{Q}_1 = \mathbb{P}[\kappa = 1]$ is the probability of the typical user served by the macrocell tier, which can be approximated by
  \begin{multline}\label{eqn:prob_1stTier}
    \mathcal{Q}_1 \approx 1 - \exp(- \pi \lambda_1 D^2) + \\
    \frac{\lambda_1}{\lambda_1 + \lambda_2 ({P_2}/{P_1})^{2/\alpha}} \cdot \exp\Big(- \pi [\lambda_1 + \lambda_2 \big(\frac{P_2}{P_1}\big)^{\frac{2}{\alpha}}] D^2 \Big),
  \end{multline}
and $\mathcal{Q}_{2,outer} = \mathbb{P}[\kappa = 2 \mid o \in A_{outer}]$ is the probability of the outer region typical user associated with the small cell tier, which can be estimated by
  \begin{align}\label{eqn:prob_2ndTier_outerarea}
    \mathcal{Q}_{2,outer} & \approx 1 - \frac{\lambda_1}{\lambda_1 + \lambda_2 ({P_2}/{P_1})^{\frac{2}{\alpha}}} \exp\Big(- \pi \lambda_2 \big(\frac{P_2}{P_1}\big)^{\frac{2}{\alpha}} D^2 \Big).
  \end{align}
\end{lemma}

\begin{IEEEproof}
See Appendix~\ref{appendix:proof_lemma_active_density}
\end{IEEEproof}

It should be noticed that $\lambda'_1$ is the density of $\Phi'_1$ over the whole plane, and $\lambda'_2$ is the density of $\Phi'_2$ in the outer region only. This difference comes from the non-uniform SCN deployment scheme.
As mentioned above, we use the homogenous PPP with density $\lambda'_1$ over the entire plane and the PPP with density $\lambda'_2$ on the outer region, respectively, to analyze the interference signal.

\subsection{Coverage Probability} \label{subsec:covprob}
Now we present the result on the coverage probability $p_{c}(T)$, that is, the probability that the instantaneous received SINR at the typical user's data channel is above a target SINR threshold $T$.\ignore{, which is equivalent to the CCDF of SINR.} The coverage probabilities provided that the typical user is located in the inner region and the outer region, are presented in Theorem~\ref{theorem:probcov_inner} and Theorem~\ref{theorem:probcov_outer}, respectively. Note that $\rho(\cdot,\cdot)$ is defined below the equation~(\ref{eqn:probcov_unifemto}).

\begin{theorem} \label{theorem:probcov_inner}
The coverage probability for the typical user in the inner region $A_{inner}$ is approximated by
  \begin{multline}\label{eqn:probcov_inner}
    p_{c, A_{inner}} (T)
    \approx \frac{2 \pi \lambda_1  }{[1 - \exp(- \pi \lambda_1 D^2)]} \int_{0}^{D}  \\
    \exp \big(- \frac{T \sigma^2 x^{\alpha}}{P_1}\big) \exp \Big(-{\pi} \big[\lambda_1 + \lambda'_1 \rho(T,\alpha)\big] {x^2} \Big) \\
    \exp \Big(-\pi \lambda'_2 D^2 \rho\big(\frac{P_2 T x^{\alpha}}{P_1 D^{\alpha}},\alpha \big) \Big) x \mathrm{d}x.
  \end{multline}
\end{theorem}

\begin{IEEEproof}
See Appendix~\ref{appendix:proof_prop_probcov_inner}.
\end{IEEEproof}

\begin{theorem} \label{theorem:probcov_outer}
The coverage probability for the typical user in the outer region $A_{outer}$ is provided by
  \begin{align}\label{eqn:probcov_outer}
    p_{c, A_{outer}} (T)
    = \sum_{i \in \{1,2\}} p_{c,i,A_{outer}} (T) \cdot \mathcal{Q}_{i,outer},
  \end{align}
where $\mathcal{Q}_{1,outer} = 1 - \mathcal{Q}_{2,outer}$, $\mathcal{Q}_{2,outer}$ is provided in Lemma~\ref{lemma:active_density}, and
  \begin{multline}\label{eqn:probcov_1stTier_outer}
    p_{c,1,A_{outer}} (T) \approx  \frac{2 \pi [\lambda_1 + \lambda_2 \big(\frac{P_2}{P_1}\big)^{2/\alpha}]}{\exp \big(- \pi [\lambda_1 + \lambda_2 \big(\frac{P_2}{P_1}\big)^{2/\alpha}] D^2 \big)} \\
    \int_{D}^{\infty} \exp \Big(- \pi \big[ \big(\lambda_1 + \lambda_2 \big(\frac{P_2}{P_1}\big)^{\frac{2}{\alpha}}\big) + \rho(T,\alpha)\big(\lambda'_1 + \lambda'_2 \big(\frac{P_2}{P_1}\big)^{\frac{2}{\alpha}}\big) \big] x^2 \Big) \\
    \exp \big(- \frac{T \sigma^2 x^{\alpha}}{P_1}\big) x \mathrm{d}x,
  \end{multline}
and
  \begin{multline}\label{eqn:probcov_2ndTier_outer}
    p_{c,2,A_{outer}} (T) = 2 \pi \lambda_2 M \cdot \int_{0}^{(\frac{P_2}{P_1})^{\frac{1}{\alpha}} D} x \exp \big(- \frac{T \sigma^2 x^{\alpha}}{P_2}\big)\\
    \exp \Big(- \pi \lambda'_1 D^2 \rho\big(\frac{P_1 T x^{\alpha}}{P_2 D^{\alpha}} , \alpha\big) - \pi \big( \lambda_2 + \lambda'_2 \rho(T,\alpha)\big) x^2 \Big)  \mathrm{d}x \\
    + \frac{2 \pi \lambda_2 M}{\exp(- \pi \lambda_1 D^2)} \int_{(\frac{P_2}{P_1})^{\frac{1}{\alpha}} D}^{\infty} x \exp \big(- \frac{T \sigma^2 x^{\alpha}}{P_2}\big) \\
    \cdot \exp \Big(- \pi \big[\big(\lambda_1 \big(\frac{P_1}{P_2}\big)^{2/\alpha} + \lambda_2\big) \\
    + \rho(T,\alpha) \big(\lambda'_1 \big(\frac{P_1}{P_2}\big)^{2/\alpha} + \lambda'_2 \big) \big] x^2 \Big) \mathrm{d}x.
  \end{multline}
\end{theorem}

\begin{IEEEproof}
See Appendix~\ref{appendix:proof_prop_probcov_outer}.
\end{IEEEproof}

The following Corollary~\ref{corol:probcov} provides the coverage probability for a randomly chosen typical user, which can be obtained easily by expanding the coverage probability into inner and outer regions.

\begin{corol} \label{corol:probcov}
For a typical user, the coverage probability is
  \begin{multline}\label{eqn:probcov_overall}
    p_{c}(T) = p_{c,A_{inner}} (T) \cdot \mathbb{P}[o \in A_{inner}] \\
    +  p_{c,A_{outer}} (T) \cdot \mathbb{P}[o \in A_{outer}],
  \end{multline}
where $\mathbb{P}[o \in A_{inner}]$ and $\mathbb{P}[o \in A_{outer}]$ are provided in (\ref{eqn:Prob_A_inner}) and~(\ref{eqn:Prob_A_outer}).
\end{corol}

\subsection{Single User Throughput}\label{subsec:single_user_throughput}
In this section, we derive another and the most important analytical result of this paper, the probabilistic characteristics of the achievable single user throughput (denoted by $\mathcal{R}$) for the proposed non-uniform SCN deployment. Similar to the coverage probability analysis in Subsection~\ref{subsec:covprob}, we will first focus on the throughput in the inner and outer regions separately. \ignore{The CCDF of the rate achieved at the typical user is also called the rate coverage \cite{SinDhi12arXiv}.}

\begin{theorem} \label{theorem:Thput_cov_innerarea}
The CCDF of the throughput achieved at the typical user in the inner region $A_{inner}$ is provided by
  \begin{multline}\label{eqn:Thput_cov_innerarea}
    \mathbb{P}[\mathcal{R} > \rho \mid o \in A_{inner}] \\
    \approx \sum_{n=0}^{\infty} \mathbb{P}[N_{1} = n] \cdot p_{c, A_{inner}} \big(2^{(n + 1)\rho /W} -1 \big),
  \end{multline}
where $N_1$ is the number of in-cell users sharing the resource with the macrocell typical user, and its PMF $\mathbb{P}[N_{1} = n]$ is approximated by
  \begin{multline}\label{eqn:PMF_MS_num_tier1}
    \mathbb{P}[N_{1} = n] \approx \frac{b^q}{n!} \cdot \frac{\Gamma(n+q+1)}{\Gamma(q)}
    \Big(\frac{\lambda_{MS}}{\lambda_1/\mathcal{Q}_{1}}\Big)^n \\
    \cdot \Big(b + \frac{\lambda_{MS}}{\lambda_1/\mathcal{Q}_{1}}\Big)^{-(n+q+1)},
    \text{ for } n \in \mathbb{Z}^0,
  \end{multline}
in which $\mathcal{Q}_{1}$ is provided in Lemma~\ref{lemma:active_density}.
\end{theorem}

\begin{IEEEproof}
See Appendix~\ref{appendix:proof_prop_Thput_cov_innerarea}.
\end{IEEEproof}

\begin{theorem} \label{theorem:Thput_cov_outerarea}
The CCDF of the throughput achieved at the typical user in the outer region $A_{outer}$ is provided by
  \begin{multline}\label{eqn:Thput_cov_outerarea}
    \mathbb{P}[\mathcal{R} > \rho \mid o \in A_{outer}] \\
    \approx \sum_{i \in \{1,2\}} \sum_{n=0}^{\infty} \mathbb{P}[N_{i} = n] p_{c, i, A_{outer}} \big(2^{(n + 1)\rho /W} -1 \big) \mathcal{Q}_{i,outer},
  \end{multline}
where $\mathcal{Q}_{1,outer} = 1 - \mathcal{Q}_{2,outer}$, $\mathcal{Q}_{2,outer}$ is provided in Lemma~\ref{lemma:active_density}, $\mathbb{P}[N_{1} = n]$ is provided in (\ref{eqn:PMF_MS_num_tier1}), and $\mathbb{P}[N_{2} = n]$ is the distribution of the number of in-cell user sharing the resource with the small cell typical user, i.e.,
  \begin{multline}\label{eqn:PMF_MS_num_tier2}
    \mathbb{P}[N_{2} = n] \approx \frac{b^q}{n!} \cdot \frac{\Gamma(n+q+1)}{\Gamma(q)} \cdot \Big(\frac{\lambda_{MS}}{\lambda_2/\mathcal{Q}_{2,outer}}\Big)^n \\
    \Big(b + \frac{\lambda_{MS}}{\lambda_2/\mathcal{Q}_{2,outer}}\Big)^{-(n+q+1)},
    \text{ for } n \in \mathbb{Z}^0.
  \end{multline}
\end{theorem}

\begin{IEEEproof}
See Appendix~\ref{appendix:proof_prop_Thput_cov_outerarea}.
\end{IEEEproof}

The following Corollary~\ref{corol:Thput_cov} provides the distribution of the throughput achieved for a typical user, which can be obtained easily by expanding the distribution expression into inner and outer regions.

\begin{corol} \label{corol:Thput_cov}
Considering resource sharing, the CCDF of the throughput achieved at the typical user can be expressed as
  \begin{multline}\label{eqn:Thput_cov}
    \mathbb{P}[\mathcal{R} > \rho] = \mathbb{P}[\mathcal{R} > \rho \mid o \in A_{inner}] \cdot \mathbb{P}[o \in A_{inner}]  \\
    + \mathbb{P}[\mathcal{R} > \rho \mid o \in A_{outer}] \cdot \mathbb{P}[o \in A_{outer}],
  \end{multline}
where $\mathbb{P}[o \in A_{inner}]$ and $\mathbb{P}[o \in A_{outer}]$ are provided in (\ref{eqn:Prob_A_inner}) and~(\ref{eqn:Prob_A_outer}).
\end{corol}


\section{Numerical Results}\label{sec:NumResults}

In this section, we present numerical results on the coverage and single user throughput for the proposed non-uniform SCN deployment scheme. Here we assume the transmit powers of macro and small cell BSs as $P_{tx,1} = 46$~dBm and $P_{tx,2} = 20$~dBm respectively. The macrocell-tier density is $\lambda_1 = 1$ per square km, and the mobile user density is $\lambda_{MS} = 10$ per square km for all numerical results. The path loss constant and exponent are assumed to be $L_0 = -34$~dB and $\alpha = 4$. The thermal noise power is $\sigma^2 = -104$~dBm. Monte Carlo simulations are also conducted to compare with our analysis for the purpose of model validation. The single user throughput demonstrated in our results are the rates achievable over the BS's bandwidth of $1$~Hz, that is, $W = 1$~Hz.

To conduct a reasonable comparison with the uniform SCN deployment with the small cell density~$\lambda_2$, we provide the numerical results for two topology scenarios in which the non-uniform SCN deployment is implemented.
\begin{itemize}
  \item In \textbf{{Scenario-I}}, all small cell BSs are uniformly deployed over the entire plane with the density $\lambda_2$; however, only the ones located in the outer region are active, while the ones in the inner region are not used. Compared with the uniform SCN deployment, the de facto density of active small cell BSs is reduced by $100\times\mathbb{P}[o \in A_{inner}]$ percent. Under this assumption, the network designer or operator can adjust the radius of the inner region to control how many small cell BSs to be used for a given network condition.
  \item In \textbf{{Scenario-II}}, we deploy all small cell BSs in the outer region, in which the small cell density is set to be $\lambda_2 / \mathbb{P}[o \in A_{outer}]$. Scenario-II guarantees that the average density over the whole plane becomes $\lambda_2$, identical to the uniform SCN deployment for a fair comparison.
\end{itemize}

\subsection{Coverage Performance}\label{subsec:CovResults}
\begin{figure}[t!]
  \centering
   \includegraphics[height=0.275\textheight, bb=70 240 520 610, clip = true]{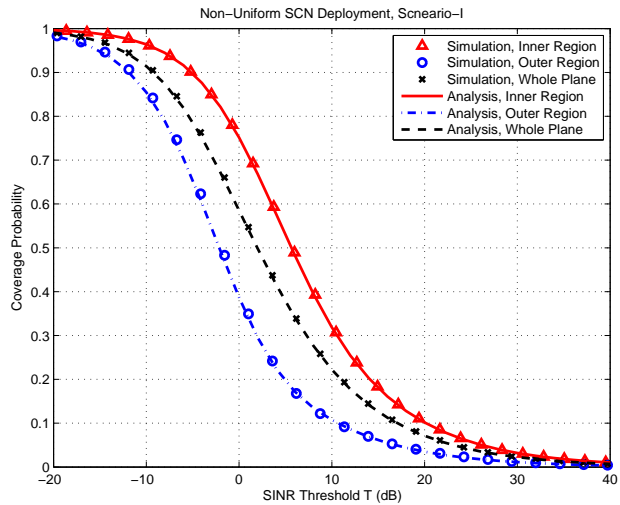}
  \caption{Coverage probability (or equivalently, the CCDF of received SINR) for Scenario-I of the proposed non-uniform SCN deployment, $D=500$~m and $\lambda_2/\lambda_1 = 10$.} \label{fig:Pc_CovOrientedDep_lambda1to10_lambdaMS10_Radius_p5}
\end{figure}

Fig.~\ref{fig:Pc_CovOrientedDep_lambda1to10_lambdaMS10_Radius_p5} demonstrates the results of coverage probability (or equivalently, the CCDF of received SINR) for Scenario-I of the proposed non-uniform SCN deployment, given the condition of $D=500$~m and $\lambda_2/\lambda_1 = 10$. Firstly, the tractable analytical results, that is, the approximations derived for inner and outer regions in this study, are reasonably accurate. Through combining the results of outer and inner regions by using (\ref{eqn:probcov_overall}), the coverage probability curve for randomly chosen users is also illustrated therein.

\begin{figure}[t!]
  \centering
  \includegraphics[height=0.275\textheight, bb=70 240 520 610, clip = true]{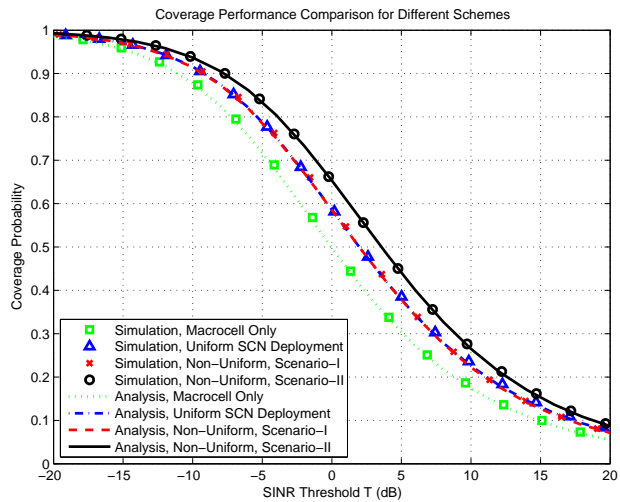}
  \caption{Coverage probability (or equivalently, the CCDF of received SINR) for different schemes, $D=500$~m and $\lambda_2/\lambda_1 = 10$.} \label{fig:Pc_WholeArea_Cmp_lambda1to10_lambdaMS10_Radius_p5}
\end{figure}

In Fig.~\ref{fig:Pc_WholeArea_Cmp_lambda1to10_lambdaMS10_Radius_p5}, the coverage probability curves for different schemes are compared, by which we can conclude that the analysis on both Scenario-I and II of the non-uniform SCN deployment precisely matches the simulation result. Furthermore, we can observe two phenomena, i.e., Scenario-I (even with reduced number of active small cell BSs) would not hurt the coverage performance, and Scenario-II outperforms both macrocell-only deployment and the uniform SCN deployment. \ignore{More importantly, users with poor reception benefit more from the smart femtocell deployment, and improving these users' performance is the key to reduce user complaints for cellular service providers.}

\begin{figure}[t!]
  \centering
  \subfigure{
    \includegraphics[height=0.275\textheight, bb=70 240 520 610, clip = true]{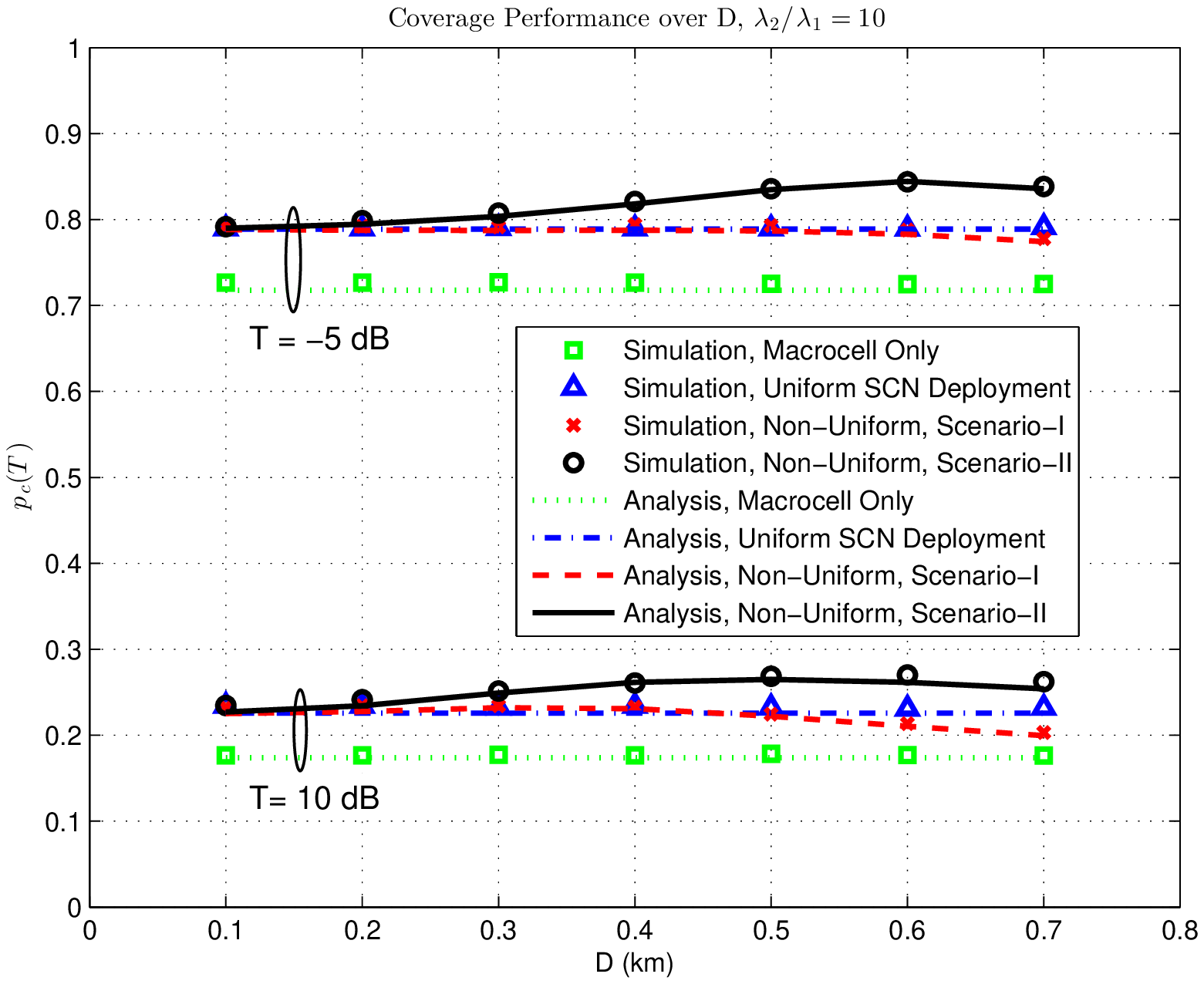}
    \label{fig:Pc_WholeArea_Cmp_lambdaMS10_overRadius:a} 
  }
  \subfigure{
    \includegraphics[height=0.275\textheight, bb=70 240 520 610, clip = true]{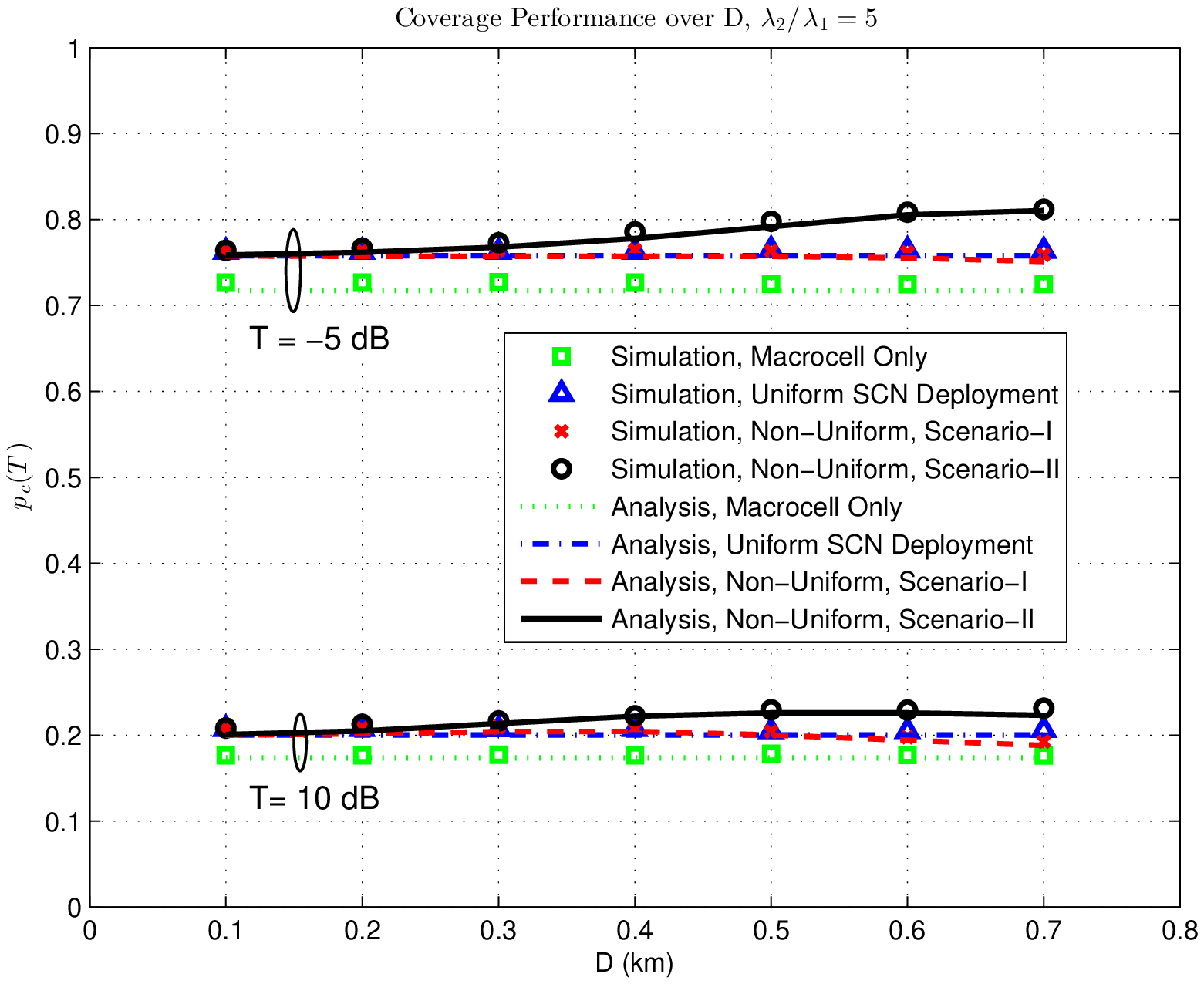}
    \label{fig:Pc_WholeArea_Cmp_lambdaMS10_overRadius:b} 
  }
  \caption{Coverage probability (or equivalently, the CCDF of received SINR) for different schemes over $D$, with the tier density ratios $\lambda_2/\lambda_1 = 10$ (upper figure) and $\lambda_2/\lambda_1 = 5$ (lower figure). The SINR thresholds are set to be $T = -5$~dB and $T = 10$~dB.} \label{fig:Pc_WholeArea_Cmp_lambdaMS10_overRadius} 
\end{figure}

Furthermore, by presenting the achievable coverage probability versus the inner region radius $D$ in Fig.~\ref{fig:Pc_WholeArea_Cmp_lambdaMS10_overRadius}, we can see the importance of properly dividing inner and outer regions on the coverage performance. For an appropriated chosen $D$ value, the proposed non-uniform SCN deployment in Scenario-I can achieve nearly the same coverage performance as the uniform case, even with a significantly lower \emph{de facto} small cell density. For instance, non-uniform Scenario-I with $D = 500$~m, which means that $54.4$\% of active small cell BSs are reduced, is as good as the uniform SCN deployment on the coverage probability for SINR threshold $T = -5$~dB. This result is surprising since more than half of the small-cell operating expense can be saved with the same level of coverage performance.

By setting the small cell density in the outer region to be $\lambda_2 / \mathbb{P}[o \in A_{outer}]$, the non-uniform SCN deployment in Scenario-II can obtain significant coverage improvement over the uniform case with the identical average small cell density. To take the case of $\lambda_2/\lambda_1 = 10$ and $T = -5$~dB as an example, the achievable coverage probability is around $85\%$ at $D = 600$~m for the non-uniform Scenario-II, compared with $79\%$ for uniform SCN deployment, and $73\%$ for single macrocell tier. Similar enhancements can be observed with a different SINR threshold (i.e., $T = 10$~dB) or a different small cell density (i.e., $\lambda_2/\lambda_1 = 5$).\ignore{By decreasing the femtocell tier density to $\lambda_2/\lambda_1 = 5$, the advantage in the coverage performance of smart femtocell deployment over the uniform deployment becomes larger, i.e., a coverage probability of $81\%$ for smart femtocell deployment and $76\%$ for uniform two-tier deployment. From both figures, similar enhancement can be observed if we consider the high performance users, i.e., $T = 10$~dB.} This benefit comes from selectively deploying SCN in the right places. It should be noticed that the slight mismatches between simulation and tractable results in Fig.~\ref{fig:Pc_WholeArea_Cmp_lambdaMS10_overRadius} come from the approximation used in the analysis, but the performance trend can be well captured by the analytical results. More importantly, both Scenario-I~and~II of the non-uniform SCN deployment do not incur any further network resources: The number of active small cell BSs is reduced in Scenario-I, and the average number of deployed small cell BSs is kept the same in Scenario-II.

\subsection{Throughput Performance}\label{subsec:TputResults}
\begin{figure}[t!]
  \centering
  \includegraphics[height=0.275\textheight, bb=65 232 520 610, clip = true]{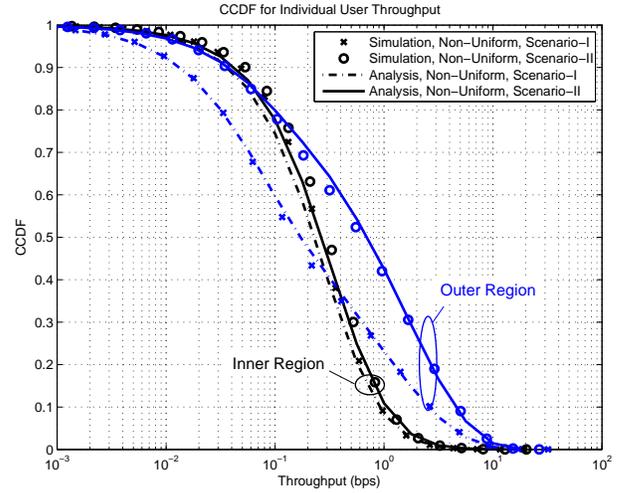}
  \caption{Single user throughput distribution (CCDF curves) for inner and outer regions, $D = 500$~m and $\lambda_2/\lambda_1 = 10$.} \label{fig:Tput_InnerOuterArea_lambda1to10_lambdaMS10_Cmp_Radius_p5}
\end{figure}

\begin{figure}[t!]
  \centering
  \includegraphics[height=0.275\textheight, bb=65 232 520 610, clip = true]{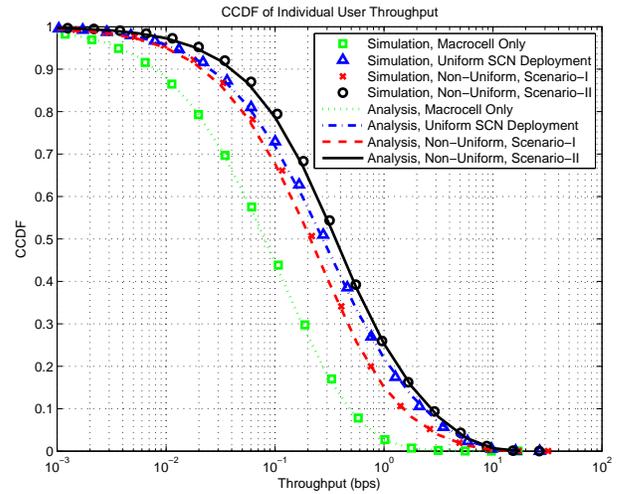}
  \caption{Single user throughput distribution (CCDF curves) for different schemes, $D = 500$~m and $\lambda_2/\lambda_1 = 10$.} \label{fig:Tput_WholeArea_Cmp_lambda1to10_lambdaMS10_Radius_p5}
\end{figure}

To validate the analytical single user throughput performances for the proposed schemes, the CCDF curves for the inner and outer regions gotten from Theorem~\ref{theorem:Thput_cov_innerarea} and Theorem~\ref{theorem:Thput_cov_outerarea} are compared with the simulation counterparts in Fig.~\ref{fig:Tput_InnerOuterArea_lambda1to10_lambdaMS10_Cmp_Radius_p5}. It can be shown that the throughput distributions are well captured by the analytical results. By combining the inner and outer region results in Corollary~\ref{corol:Thput_cov}, we compare the single user throughput performances for different deployment schemes in Fig.~\ref{fig:Tput_WholeArea_Cmp_lambda1to10_lambdaMS10_Radius_p5}. The analytical results for both Scenario-I~and~II of the non-uniform SCN deployment are reasonably accurate. For Scenario-I, basically it only reduces high-rate users' performance at the same time not hurting low-rate users, compared with the uniform SCN case. As low-rate users are usually much more of a concern to the cellular service providers \cite{EftMac06TNSPS}, this scheme has the desirable property of being able to significantly reduce the resource while taking care of the low-rate users. For Scenario-II, it increases all users performance, especially the low-rate ones. Taking the worst $10$\% users for instance, the highest achievable rate among these users is increased from $0.025$~bps in the uniform SCN case to $0.043$~bps for the proposed non-uniform SCN deployment in Scenario-II, which is a $72$\% improvement.

%

\begin{figure}[t!]
  \centering
  \subfigure{
    \includegraphics[height=0.275\textheight, bb=70 240 520 610, clip = true]{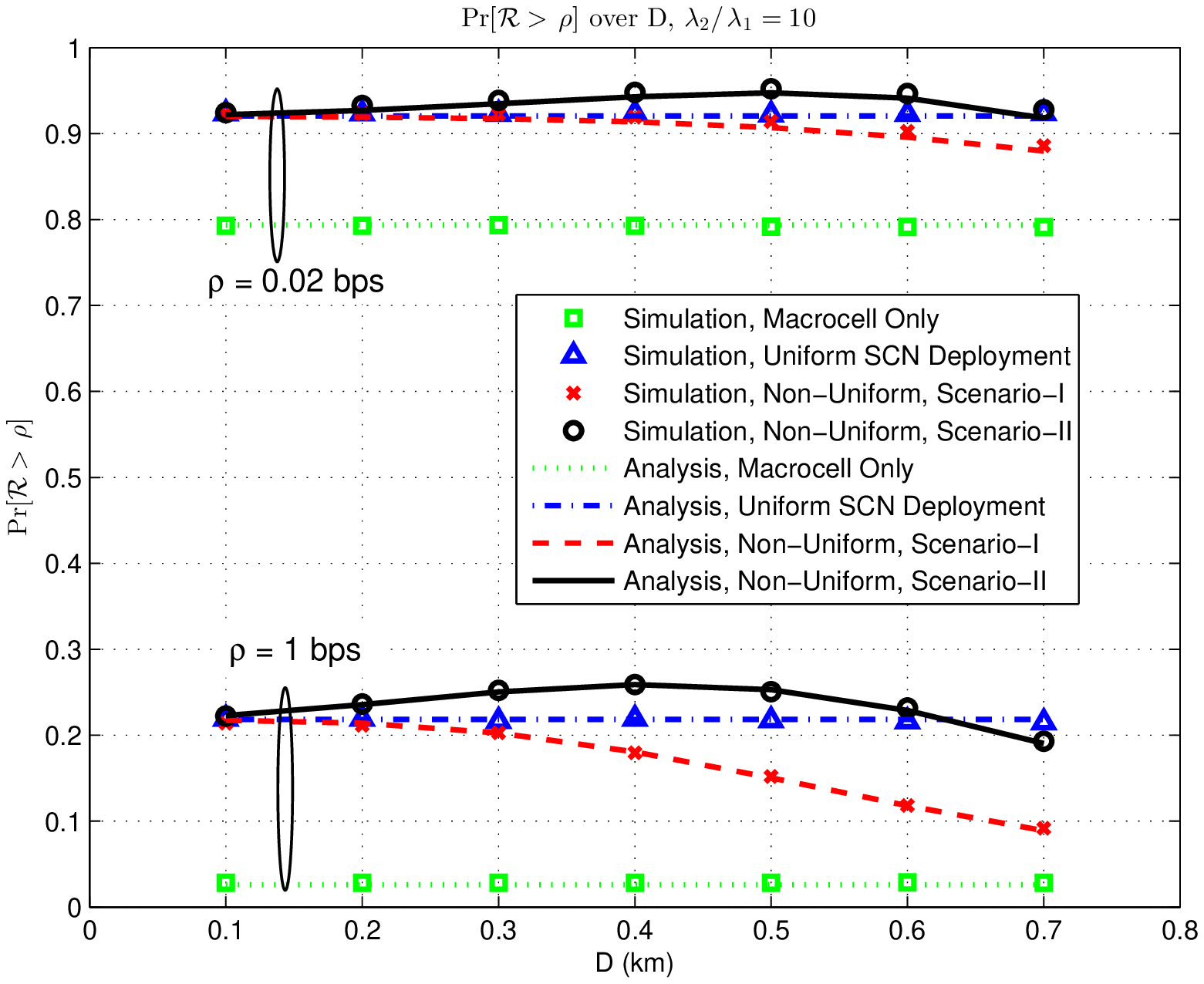}
    \label{fig:Tput_WholeArea_Cmp_lambdaMS10_overRadius:a} 
  }
  \subfigure{
    \includegraphics[height=0.275\textheight, bb=70 240 520 610, clip = true]{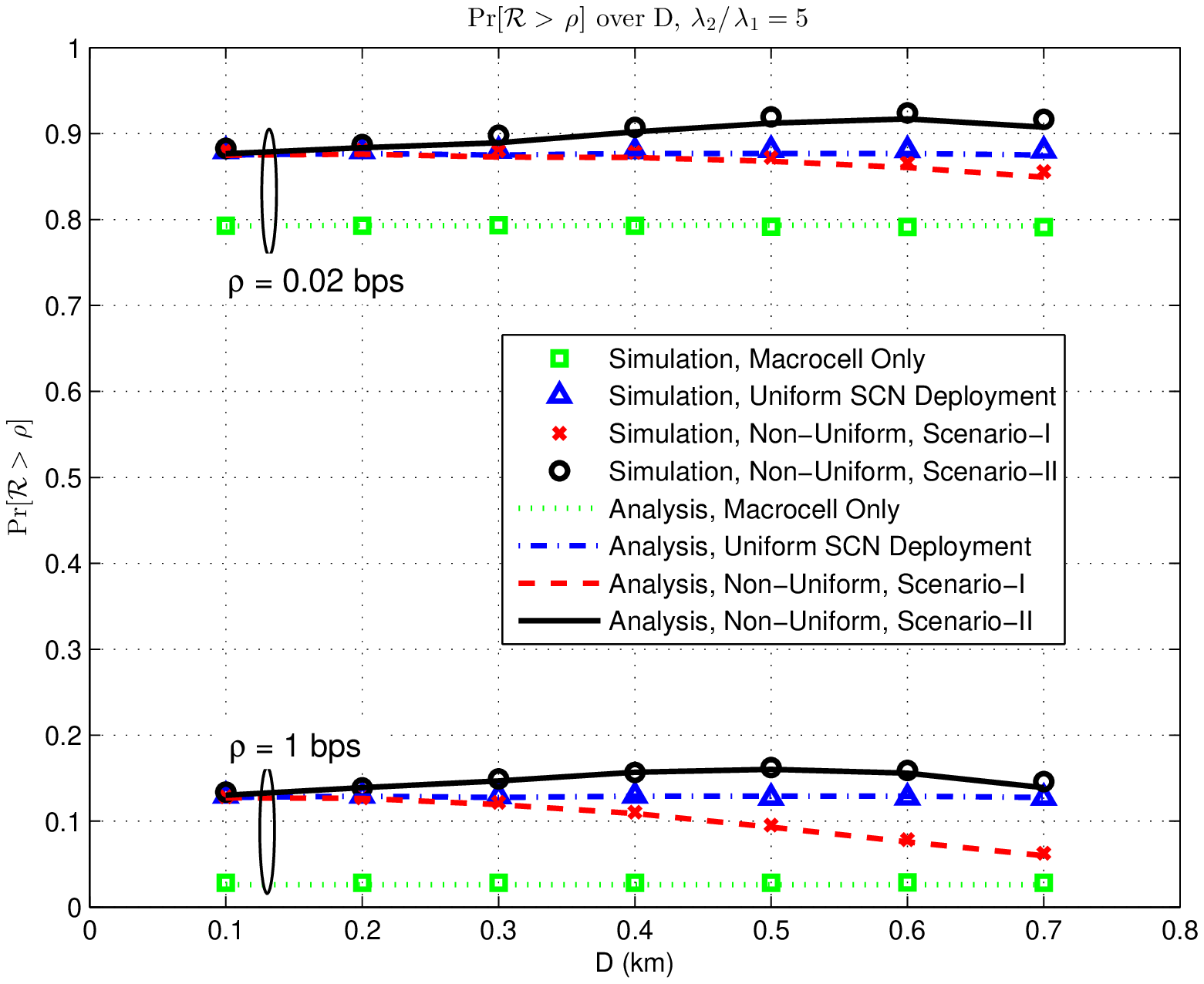}
    \label{fig:Tput_WholeArea_Cmp_lambdaMS10_overRadius:b} 
  }
  \caption{$\mathbb{P}[\mathcal{R} > \rho]$ over $D$ for different schemes, with the tier density ratios $\lambda_2/\lambda_1 = 10$ (upper figure) and $\lambda_2/\lambda_1 = 5$ (lower figure). The rate thresholds are set to be $\rho = 0.02$~bps and $\rho = 1$~bps.} \label{fig:Tput_WholeArea_Cmp_lambdaMS10_overRadius} 
\end{figure}

From the single user throughput performances versus the inner region radius $D$ demonstrated in Fig.~\ref{fig:Tput_WholeArea_Cmp_lambdaMS10_overRadius}, we can see the impact of the inner region radius $D$ on the single user throughput performance. For Scenario-I of the proposed non-uniform SCN deployment, optimally choosing $D$ can significantly reduce the resource at the same time not hurting the low-rate users' performance: for example, $D$ can get up to $400$~m in both figures for $\rho = 0.02$~bps. For Scenario-II of the non-uniform SCN deployment: optimally choosing $D$ results in noticeable improvement for both low-rate and high-rate users. Our analytical results provide tools to design the value of $D$ to maximize the benefits to a target group of users of the operator's choice. To take $\lambda_2/\lambda_1 = 10$ as an example, $D = 400$~m and $D = 500$~m can achieve near-optimal values of $\mathbb{P}[\mathcal{R} > \rho]$ for high- and low-rate thresholds, respectively. \ignore{Specifically, for high rate threshold $\rho = 1$, $\mathbb{P}[\mathcal{R} > \rho] = 26$\% is obtained for smart femtocell deployment, compared with $22$\% for the uniform deployment, which is an $18$\% increase, and a similar improvement can be observed from the case of $\lambda_2/\lambda_1 = 5$.}

\section{Conclusion}\label{sec:Conclusion}
In this study, we have studied the downlink coverage and throughput performance of the cellular networks with the newly proposed non-uniform SCN deployment scheme. Using the tools from stochastic geometry, we provided the probabilistic characterization of the downlink coverage and single user throughput at a randomly located mobile user in this new scheme. The numerical results validated the analytical expressions and approximations, and provided the following important message: By carefully choosing the parameters for the proposed non-uniform SCN deployment scheme, the active number of small cell BSs can be reduced by more than $50$\% to save operating expense, while achieving the same level of coverage performance as deploying small cells uniformly. By maintaining the average small cell density in the proposed non-uniform SCN deployment, we achieve noticeable improvement in the coverage and the data throughput ($72$\% increase for the throughput achievable by worst $10$\% users), compared with uniform deployment, with no extra cost. This interesting finding demonstrates the performance improvements achievable by implementing a simple non-uniform SCN deployment, and emphasizes the importance of deploying the small cell BSs selectively by taking their relative locations with macrocell BSs into account.



%


\appendix

\subsection{Proof of Lemma~\ref{lemma:active_density}} \label{appendix:proof_lemma_active_density}

From its definition, $\mathcal{Q}_1 = \mathbb{P}[\omega = 1]$ can be expanded as
  \begin{eqnarray}\label{eqn:prob_1stTier_derive}
    \mathcal{Q}_1 & = & \mathbb{P}[\omega = 1, o \in A_{inner}] + \mathbb{P}[\omega = 1, o \in A_{outer}] \nonumber \\
    & \approx & \mathbb{P}[o \in A_{inner}] + \mathbb{P}[\omega = 1, o \in A_{outer}],
  \end{eqnarray}
in which the approximation is based upon the fact that the inner region typical user is associated with the small cell tier with a small probability. The latter part of (\ref{eqn:prob_1stTier_derive}), i.e., the probability of accessing the macrocell tier and locating in $A_{outer}$, is provided by
  \begin{align}\label{eqn:prob_1stTier_derive2}
    &\mathbb{P}[\omega = 1, o \in A_{outer}]  \nonumber \\
    & = \mathbb{P}[\omega = 1 \mid o \in A_{outer}] \cdot \mathbb{P}[o \in A_{outer}] \nonumber \\
    & = \int_{D}^{\infty} \mathbb{P}[\omega = 1 \mid R_1 = x, o \in A_{outer}] \nonumber \\
    & \ \ \ \ \ \ \ \ \ \ \ \ \ \ \cdot f_{R_1 \mid o \in A_{outer}}(x) \mathrm{d}x \cdot \mathbb{P}[o \in A_{outer}] \nonumber \\
    & = \int_{D}^{\infty} \mathbb{P}[R_2 > \big(\frac{P_2}{P_1}\big)^{1/\alpha} x  \mid o \in A_{outer}] \nonumber \\
    & \ \ \ \ \ \ \ \ \ \ \ \ \ \ \cdot f_{R_1 \mid R_1 > D}(x) \mathrm{d}x \cdot \mathbb{P}[o \in A_{outer}] \nonumber \\
    & \stackrel{(a)}{\approx} \int_{D}^{\infty} \exp(- \pi \lambda_2 \big(\frac{P_2}{P_1}\big)^{2/\alpha} x^2 ) \cdot \frac{2 \pi \lambda_1 r \exp(- \pi \lambda_1 r^2)}{\exp(-\pi \lambda_1 D^2)} \mathrm{d}x \nonumber \\
    & \ \ \ \ \ \ \ \ \ \ \ \ \ \ \ \ \ \ \ \ \ \ \ \ \ \ \ \ \cdot \exp(-\pi \lambda_1 D^2) \nonumber \\
    & = \frac{\lambda_1}{\lambda_1 + \lambda_2 \big(\frac{P_2}{P_1}\big)^{2/\alpha}} \cdot \exp\Big(- \pi [\lambda_1 + \lambda_2 \big(\frac{P_2}{P_1}\big)^{\frac{2}{\alpha}}] D^2 \Big),
  \end{align}
where we approximate the density of small cell BSs in the vicinity of the outer region typical user as $\lambda_2$ in step $(a)$. By substituting (\ref{eqn:prob_1stTier_derive2}) and the expression of $\mathbb{P}[o \in A_{inner}]$ in (\ref{eqn:Prob_A_inner}) into (\ref{eqn:prob_1stTier_derive}), we reach the result in (\ref{eqn:prob_1stTier}).

This result $\mathcal{Q}_{2,outer} = \mathbb{P}[\omega = 2 \mid o \in A_{outer}]$ can be obtained by $\mathcal{Q}_{2,outer} = 1 - {\mathbb{P}[\omega = 1, o \in A_{outer}]}/{\mathbb{P}[o \in A_{outer}]}$, in which $\mathbb{P}[\omega = 1, o \in A_{outer}]$ and $\mathbb{P}[o \in A_{outer}]$ are provided in (\ref{eqn:prob_1stTier_derive2}) and (\ref{eqn:Prob_A_outer}).

Similar to the method used in \cite{SinDhi12arXiv}, we use the Voronoi cell area formed by a homogeneous PPP with certain density values to approximate the area of the macro and small cells. The area of the macrocell tier cells can be approximated by
  \begin{align}\label{eqn:vc_tier1}
    \mathcal{C}_{1} \approx \mathcal{C}_0 \big(\frac{\lambda_1}{\mathcal{Q}_{1}}\big).
  \end{align}
On the other hand, the SCN-deployed region $A_{outer}$ is the area where small cell BSs are deployed with the density $\lambda_2$.
Hence, the area of cells formed by the small cell tier can be similarly approximated by
  \begin{align}\label{eqn:vc_tier2}
    \mathcal{C}_{2} \approx \mathcal{C}_0 \big(\frac{\lambda_2}{\mathcal{Q}_{2,outer}}\big).
  \end{align}

Similar to the analysis in Section~\ref{sec:SysModel}, the probabilities of the $i$-th tier cells with no user associated are $\mathbb{P}[N_{i,c} = 0]$, in which $N_{i,c}$ is the number of users in a randomly chosen $i$-th tier cell,
  \begin{align}\label{eqn:PMF_MS_num_tier1_RC}
    \mathbb{P}[N_{1,c} = n] \approx \frac{b^q}{n!} \cdot \frac{\Gamma(n+q)}{\Gamma(q)} \cdot \frac{(\lambda_{MS})^n (\lambda_1/\mathcal{Q}_{1})^q}{(\lambda_{MS} + b \lambda_1/\mathcal{Q}_{1})^{n+q}},
  \end{align}
and
  \begin{align}\label{eqn:PMF_MS_num_tier2_RC}
    \mathbb{P}[N_{2,c} = n] \approx \frac{b^q}{n!} \cdot \frac{\Gamma(n+q)}{\Gamma(q)} \cdot \frac{(\lambda_{MS})^n (\lambda_2/\mathcal{Q}_{2,outer})^q}{(\lambda_{MS} + b \lambda_2/\mathcal{Q}_{2,outer})^{n+q}}.
  \end{align}
Then the average density can be obtained by $\lambda'_i = \lambda_i \cdot (1 - \mathbb{P}[N_{i,c} = 0])$, which completes the proof.

\subsection{Proof of Theorem~\ref{theorem:probcov_inner}} \label{appendix:proof_prop_probcov_inner}

We define the random variable $X_i$ as the distance between the typical user and its serving BS, given the condition that the user is served by the $i$-th tier. $R_i$ is defined as the typical user's distance from the nearest BS in the $i$-th tier. It should be noted that $R_i$ does not request the serving BS is served by the $i$-tier, which is different from $X_i$. The relationship between $X_i$ and $R_i$ is $\mathbb{P}[X_i \leqslant x] = \mathbb{P}[R_i \leqslant x \mid \kappa = i]$. Firstly, we will derive the PDFs of $X_1$ for the typical user in $A_{inner}$.

Since the event of $X_1 \leqslant x$ is the event of $R_1 \leqslant x$ based on the condition that the user is associated with macrocell, the cumulative distribution function (CDF) of $X_1$ for the inner region typical user can be expressed as
  \begin{align}\label{eqn:cdf_X1_cond_innerarea}
    F_{X_1 \mid o \in A_{inner}}(x) & = \mathbb{P}[X_1 \leqslant x \mid o \in A_{inner}] \nonumber \\
    & = \mathbb{P}[R_1 \leqslant x \mid \kappa = 1, o \in A_{inner}] \nonumber \\
    & \stackrel{(a)}{\approx} \mathbb{P}[R_1 \leqslant x \mid R_1 \leqslant D] \nonumber \\
    & = \frac{1 - \exp(- \pi \lambda_1 x^2)}{1 - \exp(- \pi \lambda_1 D^2)}, \ \ \ \ \ \text{for $x \leqslant D$},
  \end{align}
where the approximation in step $(a)$ is conducted by assuming that the inner region typical user always gets service from the macrocell tier. This assumption is reasonable from the practical implementation viewpoint: The transmit powers of macrocell BSs should be much larger than small cell BSs, that is, $P_1 \gg P_2$, which makes the fact that the inner region typical user is served by a small cell BS with a small probability.

Subsequently, the PDF of $X_1$ can be found by differentiating the CDF, i.e.,
  \begin{multline}\label{eqn:pdf_X1_cond_innerarea}
    f_{X_1 \mid o \in A_{inner}}(x) = \frac{\mathrm{d} F_{X_1 \mid o \in A_{inner}}(x)} {\mathrm{d} x}   \\
    \approx  \frac{2 \pi \lambda_1 x  }{[1 - \exp(- \pi \lambda_1 D^2)]} \exp(- \pi \lambda_1 x^2), \ \text{for $x \leqslant D$}.
  \end{multline}

By assuming that the typical user located in the inner region $A_{inner}$ always gets service from macrocell BSs, its coverage probability can be derived as
  \begin{align}\label{eqn:probcov_inner_derive}
    & p_{c, A_{inner}} (T) 
    \approx \mathbb{P}[\mathrm{SINR}_1 > T \mid o \in A_{inner}] \nonumber \\
    & = \int_{0}^{D} \mathbb{P}[\mathrm{SINR}_1 > T \mid X_1 = x, o \in A_{inner}] \nonumber \\
    & \ \ \ \ \ \ \ \ \ \ \ \ \ \ \ \ \ \ \ \ \ \ \ \ \ \ \ \ \ \ \ \ \ \ \cdot f_{X_1 \mid o \in A_{inner}}(x) \mathrm{d}x \nonumber \\
    & \stackrel{(a)}{=} \int_{0}^{D} \exp \big(- \frac{T \sigma^2 x^{\alpha}}{P_1}\big) \prod_{i=1}^2 \mathcal{L}_{I_i \mid X_1 = x, o \in A_{inner}}\big(\frac{T x^{\alpha}}{P_1} \big) \nonumber \\
    & \ \ \ \ \ \ \ \ \ \ \ \ \ \ \ \ \ \ \ \ \ \ \ \ \ \ \ \ \ \ \ \ \ \ \cdot f_{X_1 \mid o \in A_{inner}}(x) \mathrm{d}x,
  \end{align}
where step $(a)$ comes from the Rayleigh fading assumption, and $\mathcal{L}_{I_i \mid X_1 = x, o \in A_{inner}}( \cdot )$ is the Laplace transform of random variable $I_i$ given the condition that the typical user $x$ away from the macrocell serving BS is located in the inner region $A_{inner}$. Here we assume the small cell interference comes from the whole region out of the area $B(o,D)$, which is an optimistic estimation (proved to be accurate by the numerical results). Then, we can have
  \begin{multline}\label{eqn:laplacetransform_inner}
  \prod_{i=1}^2 \mathcal{L}_{I_i \mid X_1 = x, o \in A_{inner}}\big(\frac{T x^{\alpha}}{P_1} \big) \\
  \approx \exp \big(-{\pi} \lambda'_1 \rho(T,\alpha) {x^2} \big) \exp \big(-\pi \lambda'_2 D^2 \rho(\frac{P_2 T x^{\alpha}}{P_1 D^{\alpha}},\alpha) \big).
  \end{multline}
By substituting (\ref{eqn:laplacetransform_inner}) and (\ref{eqn:pdf_X1_cond_innerarea}) into (\ref{eqn:probcov_inner_derive}), we can obtain (\ref{eqn:probcov_inner}) and complete the proof.

\subsection{Proof of Theorem~\ref{theorem:probcov_outer}} \label{appendix:proof_prop_probcov_outer}
Because the event of $X_1 \leqslant x$ is the event of $R_1 \leqslant x$ provided that the user is associated with macrocell, the CDF of $X_1$ for the outer region typical user is
  \begin{align}\label{eqn:cdf_X1_cond_outerarea}
    & F_{X_1 \mid o \in A_{outer}}(x) \nonumber \\
    & = \mathbb{P}[X_1 \leqslant x \mid o \in A_{outer}] \nonumber \\
    & = \mathbb{P}[R_1 \leqslant x \mid \kappa = 1, o \in A_{outer}] \nonumber \\
    & = \mathbb{P}[R_1 \leqslant x \mid R_1 < \big(\frac{P_1}{P_2}\big)^{1/\alpha} R_2, R_1 > D] \nonumber \\
    & \stackrel{(a)}{=} \frac{\mathbb{P}[R_1 \leqslant x, R_1 < \big(\frac{P_1}{P_2}\big)^{1/\alpha} R_2 \mid R_1 > D] \cdot \mathbb{P}[R_1 > D]}{\mathbb{P}[R_1 < \big(\frac{P_1}{P_2}\big)^{1/\alpha} R_2, R_1 > D]}, \nonumber \\
    & \ \ \ \ \ \ \ \ \ \ \ \ \ \ \ \ \ \ \ \ \ \ \ \ \ \ \ \ \ \ \ \ \ \ \ \ \ \ \ \ \ \ \ \ \ \ \text{for $x > D$},
  \end{align}
in which step $(a)$ follows Bayes' theorem. The denominator of (\ref{eqn:cdf_X1_cond_outerarea}) can be derived as
  \begin{align}\label{eqn:cdf_X1_cond_outerarea_derive1}
    \mathbb{P}&[R_1 < \big(\frac{P_1}{P_2}\big)^{1/\alpha} R_2, R_1 > D] \nonumber \\
    & = \int_{D (\frac{P_2}{P_1})^{1/\alpha}}^{\infty} \mathbb{P}[D < R_1 < \big(\frac{P_1}{P_2}\big)^{1/\alpha} r] f_{R_2}(r) \mathrm{d}r \nonumber \\
    & \stackrel{(b)}{\approx} \int_{D (\frac{P_2}{P_1})^{1/\alpha}}^{\infty} \big[\exp(-\pi \lambda_1 D^2) - \exp(- \pi \lambda_1 \big(\frac{P_1}{P_2}\big)^{2/\alpha} r^2) \big] \nonumber \\
    & \ \ \ \ \ \ \ \ \ \ \ \ \ \cdot 2 \pi \lambda_2 r \exp(- \pi \lambda_2 r^2) \mathrm{d}r \nonumber \\
    & = \frac{\lambda_1}{\lambda_1 + \lambda_2 \big(\frac{P_2}{P_1}\big)^{2/\alpha}} \cdot \exp \big(- \pi [\lambda_1 + \lambda_2 \big(\frac{P_2}{P_1}\big)^{2/\alpha}] D^2 \big),
  \end{align}
where we approximate the density of small cell BSs in the vicinity of the outer region typical user as $\lambda_2$, and follow the PPP's void probability in step $(b)$. This approximation is accurate as long as the typical user is not located close to the boundary between the inner and outer regions. Based upon the same approximation, the former part of (\ref{eqn:cdf_X1_cond_outerarea})'s numerator is derived~as
  \begin{align}\label{eqn:cdf_X1_cond_outerarea_derive2}
    \mathbb{P}&[R_1 \leqslant x, R_1 < \big(\frac{P_1}{P_2}\big)^{1/\alpha} R_2 \mid R_1 > D] \nonumber \\
    \ignore{& = & \int_D^{\infty} \mathbb{P}[r \leqslant x, R_2 > \big(\frac{P_2}{P_1}\big)^{1/\alpha} r] f_{R_1 \mid R_1 >D}(r) \mathrm{d}r \nonumber \\}
    & = \int_D^{x} \mathbb{P}[R_2 > \big(\frac{P_2}{P_1}\big)^{1/\alpha} r] f_{R_1 \mid R_1 >D}(r) \mathrm{d}r \nonumber \\
    & \approx \int_D^{x} \exp(- \pi \lambda_2 \big(\frac{P_2}{P_1}\big)^{2/\alpha} r^2) \cdot \frac{2 \pi \lambda_1 r \exp(- \pi \lambda_1 r^2)}{\exp(- \pi \lambda_1 D^2)} \mathrm{d}r \nonumber \\
    & = \frac{\lambda_1}{\lambda_1 + \lambda_2 \big(\frac{P_2}{P_1}\big)^{2/\alpha}} \cdot \frac{1}{\exp(- \pi \lambda_1 D^2)} \nonumber \\
    & \ \ \ \ \ \ \ \ \ \ \ \cdot \Big[\exp \big(- \pi [\lambda_1 + \lambda_2 \big(\frac{P_2}{P_1}\big)^{2/\alpha}] D^2 \big) \nonumber \\
    & \ \ \ \ \ \ \ \ \ \ \ \ \ \ \ \ \ - \exp \big(- \pi [\lambda_1 + \lambda_2 \big(\frac{P_2}{P_1}\big)^{2/\alpha}] x^2 \big) \Big],
  \end{align}
in which similar techniques are applied as we derive (\ref{eqn:cdf_X1_cond_outerarea_derive1}).

By substituting (\ref{eqn:cdf_X1_cond_outerarea_derive1}) and (\ref{eqn:cdf_X1_cond_outerarea_derive2}) into (\ref{eqn:cdf_X1_cond_outerarea}) and differentiating the resultant CDF, we can reach $X_1$'s PDF, that is,
  \begin{multline}\label{eqn:pdf_X1_cond_outerarea}
    f_{X_1 \mid o \in A_{outer}}(x) \approx \\
    \frac{2 \pi [\lambda_1 + \lambda_2 \big(\frac{P_2}{P_1}\big)^{\frac{2}{\alpha}}] \cdot x \exp \big(- \pi [\lambda_1 + \lambda_2 \big(\frac{P_2}{P_1}\big)^{\frac{2}{\alpha}}] x^2 \big)}{\exp \big(- \pi [\lambda_1 + \lambda_2 \big(\frac{P_2}{P_1}\big)^{\frac{2}{\alpha}}] D^2 \big)} , \\
    \ \ \text{for $x > D$}.
  \end{multline}

Because the event of $X_2 \leqslant x$ is the event of $R_2 \leqslant x$ based on the condition that the user is associated with the small cell tier, the CDF of $X_2$ for the outer region typical user can be derived as
  \begin{align}\label{eqn:cdf_X2_cond_outerarea}
    & F_{X_2 \mid o \in A_{outer}}(x) \nonumber \\
    & = \mathbb{P}[X_2 \leqslant x \mid o \in A_{outer}] \nonumber \\
    & = \mathbb{P}[R_2 \leqslant x \mid \kappa = 2, o \in A_{outer}] \nonumber \\
    & = \mathbb{P}[R_2 \leqslant x \mid R_2 < \big(\frac{P_2}{P_1}\big)^{1/\alpha} R_1, R_1 > D] \nonumber \\
    & \stackrel{(a)}{=} \frac{\mathbb{P}[R_2 \leqslant x, R_2 < \big(\frac{P_2}{P_1}\big)^{1/\alpha} R_1 \mid R_1 > D] \cdot \mathbb{P}[R_1 > D]}{\mathbb{P}[R_2 < \big(\frac{P_2}{P_1}\big)^{1/\alpha} R_1, R_1 > D]}, \nonumber \\
    & \ \ \ \ \ \ \ \ \ \ \ \ \ \ \ \ \ \ \ \ \ \ \ \ \ \ \ \ \ \ \ \ \ \ \ \ \ \ \ \ \ \ \ \ \ \ \text{for $x > D$},
  \end{align}
where step $(a)$ follows Bayes' theorem. The former part of (\ref{eqn:cdf_X2_cond_outerarea})'s numerator is expressed as
  \begin{eqnarray}\label{eqn:cdf_X2_cond_outerarea_derive2}
    & & \mathbb{P}[R_2 \leqslant x, R_2 < \big(\frac{P_2}{P_1}\big)^{1/\alpha} R_1 \mid R_1 > D] \nonumber \\
    & = & \begin{cases}
      \mathbb{P}[R_2 \leqslant x] \ \ \ \ \ \ \ \ \ \ \ \ \ \ \ \ \ \ \ \ \ \text{for $x \leqslant \big(\frac{P_2}{P_1}\big)^{1/\alpha} D $}\\
      \int_D^{\infty} \mathbb{P}[R_2 \leqslant x, R_2 < \big(\frac{P_2}{P_1}\big)^{1/\alpha} r] f_{R_1 \mid R_1 > D}(r) \mathrm{d}r \\
      \ \ \ \ \ \ \ \ \ \ \ \ \ \ \ \ \ \ \ \ \ \ \ \ \ \ \ \ \ \ \ \ \ \text{for $x > \big(\frac{P_2}{P_1}\big)^{1/\alpha} D $},\\
    \end{cases} \nonumber \\
  \end{eqnarray}
in which the integration under the condition of $x > ({P_2}/{P_1})^{1/\alpha} D$ can be derived as
  \begin{align}\label{eqn:cdf_X2_cond_outerarea_derive3}
    & \int_D^{\infty} \mathbb{P}[R_2 \leqslant x, R_2 < \big(\frac{P_2}{P_1}\big)^{1/\alpha} r] f_{R_1 \mid R_1 > D}(r) \mathrm{d}r \nonumber \\
    & = \int_{x (\frac{P_1}{P_2})^{1/\alpha}}^{\infty} \mathbb{P}[R_2 < x] \cdot \frac{2 \pi \lambda_1 r \exp(- \pi \lambda_1 r^2)}{\exp(-\pi \lambda_1 D^2)} \mathrm{d}r \nonumber \\
    & \ + \int_D^{x (\frac{P_1}{P_2})^{1/\alpha} } \mathbb{P}[R_2 < \big(\frac{P_2}{P_1}\big)^{1/\alpha} r] \cdot \frac{2 \pi \lambda_1 r \exp(- \pi \lambda_1 r^2)}{\exp(-\pi \lambda_1 D^2)} \mathrm{d}r \nonumber \\
    & \stackrel{(b)}{\approx} \int_{x (\frac{P_1}{P_2})^{1/\alpha}}^{\infty} \big[1 - \exp(-\pi \lambda_2 x^2)\big] \cdot \frac{2 \pi \lambda_1 r \exp(- \pi \lambda_1 r^2)}{\exp(-\pi \lambda_1 D^2)} \mathrm{d}r \nonumber \\
    & \ \ \ \ + \int_D^{x (\frac{P_1}{P_2})^{1/\alpha} } \big[1 - \exp(-\pi \lambda_2 \big(\frac{P_2}{P_1}\big)^{2/\alpha} r^2)\big] \nonumber \\
    & \ \ \ \ \ \ \ \ \ \ \ \ \ \ \ \ \ \ \ \ \ \ \ \ \ \ \ \ \ \ \ \ \ \ \ \cdot \frac{2 \pi \lambda_1 r \exp(- \pi \lambda_1 r^2)}{\exp(-\pi \lambda_1 D^2)} \mathrm{d}r \nonumber \\
    & = 1 - \frac{1}{\exp(-\pi \lambda_1 D^2)} \cdot \frac{\lambda_1}{\lambda_1 + \lambda_2 \big(\frac{P_2}{P_1}\big)^{2/\alpha}} \nonumber \\
    & \ \ \ \ \ \ \ \ \ \ \ \ \ \ \ \ \ \ \ \ \ \ \ \ \ \ \cdot \exp \big(- \pi [\lambda_1 + \lambda_2 \big(\frac{P_2}{P_1}\big)^{2/\alpha}] D^2 \big) \nonumber \\
    & \ \ \ \ - \frac{1}{\exp(-\pi \lambda_1 D^2)} \cdot \frac{\lambda_2 \big(\frac{P_2}{P_1}\big)^{2/\alpha}}{\lambda_1 + \lambda_2 \big(\frac{P_2}{P_1}\big)^{2/\alpha}} \nonumber \\
    & \ \ \ \ \ \ \ \ \ \ \ \ \ \ \ \ \ \ \ \ \ \ \cdot \exp \big(- \pi [\lambda_1 \big(\frac{P_1}{P_2}\big)^{2/\alpha} + \lambda_2] x^2 \big),
  \end{align}
in which step $(b)$ is approximated by assuming that the density of small cell BSs in the vicinity of the outer region typical user is $\lambda_2$. The same approximation will help us to derive the denominator of (\ref{eqn:cdf_X2_cond_outerarea}), i.e.,
  \begin{align}\label{eqn:cdf_X2_cond_outerarea_derive1}
    \mathbb{P}&[R_2 < \big(\frac{P_2}{P_1}\big)^{1/\alpha} R_1, R_1 > D] \nonumber \\
    & = \int_D^{\infty} \mathbb{P}[R_2 < \big(\frac{P_2}{P_1}\big)^{1/\alpha} r] \cdot 2 \pi \lambda_1 r \exp(- \pi \lambda_1 r^2) \mathrm{d}r \nonumber \\
    & \approx \int_D^{\infty} \big[1 - \exp(- \pi \lambda_2 \big(\frac{P_2}{P_1}\big)^{\frac{2}{\alpha}} r^2)\big] 2 \pi \lambda_1 r \exp(- \pi \lambda_1 r^2) \mathrm{d}r \nonumber \\
    & = \exp(- \pi \lambda_1 D^2) - \frac{\lambda_1}{\lambda_1 + \lambda_2 \big(\frac{P_2}{P_1}\big)^{2/\alpha}} \nonumber \\
    & \ \ \ \ \ \ \ \ \ \ \ \ \ \ \ \ \ \ \ \ \ \cdot \exp \big(- \pi [\lambda_1 + \lambda_2 \big(\frac{P_2}{P_1}\big)^{2/\alpha}] D^2 \big).
  \end{align}
By substituting (\ref{eqn:cdf_X2_cond_outerarea_derive1}), (\ref{eqn:cdf_X2_cond_outerarea_derive2}) and (\ref{eqn:cdf_X2_cond_outerarea_derive3}) into (\ref{eqn:cdf_X2_cond_outerarea}) and differentiating the resultant CDF, $X_2$'s PDF can be obtained as below,
  \begin{align}\label{eqn:pdf_X2_cond_outerarea}
    & f_{X_2 \mid o \in A_{outer}}(x) \nonumber \\
    & \approx \begin{cases}
      M \cdot 2 \pi \lambda_2 x \exp(- \pi \lambda_2 x^2),\ \ \ \ \ \ \ \ \ \ \text{for $x \leqslant \big(\frac{P_2}{P_1}\big)^{1/\alpha} D $}\\
      M \cdot \Big[\frac{2 \pi x \lambda_2}{\exp(- \pi \lambda_1 D^2)} \cdot \exp \big(- \pi [\lambda_1 \big(\frac{P_1}{P_2}\big)^{2/\alpha} + \lambda_2] x^2 \big) \Big], \\
      \ \ \ \ \ \ \ \ \ \ \ \ \ \ \ \ \ \ \ \ \ \ \ \ \ \ \ \ \ \ \ \ \ \ \ \ \ \ \ \ \ \text{for $x > \big(\frac{P_2}{P_1}\big)^{1/\alpha} D $}, \\
    \end{cases} \nonumber \\
    &
  \end{align}
where the constant $M$ is
  \begin{multline}\label{eqn:M_constant}
    M = {\exp(-\pi \lambda_1 D^2)}/ \Big[ \exp(-\pi \lambda_1 D^2) - \\
    \lambda_1 \exp \big(- \pi [\lambda_1 + \lambda_2 (\frac{P_2}{P_1})^{2/\alpha}] D^2 \big)/(\lambda_1 + \lambda_2 (\frac{P_2}{P_1})^{2/\alpha}) \Big].
  \end{multline}

For the outer region typical user served by the macrocell tier, its coverage probability can be expressed as
  \begin{align}\label{eqn:probcov_1stTier_outer_derive}
    & p_{c,1,A_{outer}} (T) = \mathbb{P}[\mathrm{SINR}_1 > T \mid o \in A_{outer}] \nonumber \\
    & = \int_{D}^{\infty} \mathbb{P}[\mathrm{SINR}_1 > T \mid X_1 = x, o \in A_{outer}] \nonumber \\
    & \ \ \ \ \ \ \ \ \ \ \ \ \ \ \ \ \ \ \ \ \ \ \ \ \ \ \ \ \ \ \ \ \ \ \ f_{X_1 \mid o \in A_{outer}}(x) \mathrm{d}x \nonumber \\
    & \stackrel{(a)}{=} \int_{D}^{\infty} \exp \big(- \frac{T \sigma^2 x^{\alpha}}{P_1}\big) \nonumber \\
    & \ \ \ \cdot \prod_{i=1}^2 \mathcal{L}_{I_i \mid X_1 = x, o \in A_{outer}}\big(\frac{T x^{\alpha}}{P_1} \big) f_{X_1 \mid o \in A_{outer}}(x) \mathrm{d}x,
  \end{align}
where step $(a)$ still follows from the Rayleigh fading assumption, and ${\mathcal{L}}_{I_{i} \mid X_j = x, o \in A_{outer}} (\cdot)$ is the Laplace transform of random variable $I_i$ given the condition that the typical user $x$ away from the macrocell serving BS is located in the outer region $A_{outer}$. Assuming the interference from the small cell tier comes from the whole plane, we can approximate this Laplace transform as
  \begin{multline}\label{eqn:laplacetransform_1stTier_outer}
    \mathcal{L}_{I_i \mid X_1 = x, o \in A_{outer}}\big(\frac{T x^{\alpha}}{P_1} \big) \\
    \approx \exp \big(- \pi x^2 \rho(T,\alpha) \lambda'_i \big(\frac{P_i}{P_1}\big)^{2/\alpha} \big).
  \end{multline}
It is a pessimistic assumption since the original small cell interference from inner area is eliminated due to this non-uniform SCN deployment scheme, and the numerical results in Section~\ref{sec:NumResults} show that it is still a reasonably accurate approximation.

By substituting the PDF of $X_1$ conditioned on that the typical user is in the outer region and provided in (\ref{eqn:pdf_X1_cond_outerarea}), and the result in (\ref{eqn:laplacetransform_1stTier_outer}) into (\ref{eqn:probcov_1stTier_outer_derive}), we can have
  \begin{align}\label{eqn:probcov_1stTier_outer_derive_2}
    & p_{c,1,A_{outer}} (T) \nonumber \\
    & \approx \int_{D}^{\infty} 2 \pi \exp \big(- \frac{T \sigma^2 x^{\alpha}}{P_1}\big) \exp \big(- \pi [\lambda_1 + \lambda_2 \big(\frac{P_2}{P_1}\big)^{\frac{2}{\alpha}}] x^2 \big) \nonumber \\
    & \exp \big(- \pi x^2 \rho(T,\alpha) [\lambda'_1 + \lambda'_2 \big(\frac{P_2}{P_1}\big)^{2/\alpha}] \big) [\lambda_1 + \lambda_2 \big(\frac{P_2}{P_1}\big)^{2/\alpha}]\nonumber \\
    & \ \ \ \ \ \ \ \ \ \ \ \ \ \ \ \ \ \ \cdot x /\Big[\exp \big(- \pi [\lambda_1 + \lambda_2 \big(\frac{P_2}{P_1}\big)^{2/\alpha}] D^2 \big) \Big] \mathrm{d}x \nonumber \\
    & = \frac{2 \pi [\lambda_1 + \lambda_2 \big(\frac{P_2}{P_1}\big)^{2/\alpha}]}{\exp \big(- \pi [\lambda_1 + \lambda_2 \big(\frac{P_2}{P_1}\big)^{2/\alpha}] D^2 \big)} \int_{D}^{\infty} \exp \big(- \frac{T \sigma^2 x^{\alpha}}{P_1}\big) \nonumber \\
    & \ \ \ \ \ \ \ \ \ \ \ \ \ \ \ \ \cdot \exp \Big(- \pi \big[ \big(\lambda_1 + \lambda_2 \big(\frac{P_2}{P_1}\big)^{2/\alpha}\big) + \nonumber \\
    & \ \ \ \ \ \ \ \ \ \ \ \ \ \ \ \ \ \ \ \ \ \rho(T,\alpha)\big( \lambda'_1 + \lambda'_2 \big(\frac{P_2}{P_1}\big)^{2/\alpha}\big) \big] x^2 \Big) x \mathrm{d}x.
  \end{align}
which completes the proof for the result in (\ref{eqn:probcov_1stTier_outer}).

If the outer region typical user is served by the small cell tier, its coverage probability can be expressed as
  \begin{align}\label{eqn:probcov_2ndTier_outer_derive}
    & p_{c,2,A_{outer}} (T) = \mathbb{P}[\mathrm{SINR}_2 > T \mid o \in A_{outer}] \nonumber \\
    & = \int_{0}^{\infty} \mathbb{P}[\mathrm{SINR}_2 > T \mid X_2 = x, o \in A_{outer}] \nonumber \\
    & \ \ \ \ \ \ \ \ \ \ \ \ \ \ \ \ \ \ \ \ \ \ \ \ \ \ \ \ \ \ \ \ \ \ \ \ \cdot f_{X_2 \mid o \in A_{outer}}(x) \mathrm{d}x \nonumber \\
    & \stackrel{(a)}{=} \int_{0}^{\infty} \exp \big(- \frac{T \sigma^2 x^{\alpha}}{P_2}\big) \nonumber \\
    & \ \ \cdot \prod_{i=1}^2 \mathcal{L}_{I_i \mid X_2 = x, o \in A_{outer}}\big(\frac{T x^{\alpha}}{P_2} \big) f_{X_2 \mid o \in A_{outer}}(x) \mathrm{d}x,
  \end{align}
where step $(a)$ follows from the Rayleigh fading assumption. The Laplace transform of random variable $I_2$ given the condition that the typical user $x$ away from the serving small cell BS is located in the outer region $A_{outer}$ is 
  \begin{eqnarray}\label{eqn:laplacetransform_2ndTier_outer}
    \mathcal{L}_{I_2 \mid X_2 = x, o \in A_{outer}}\big(\frac{T x^{\alpha}}{P_2} \big) \approx \exp \big(- \pi \lambda'_2 \rho(T,\alpha) x^2 \big),
  \end{eqnarray}
and the Laplace transform of random variable $I_1$ given that condition can be derived as
  \begin{align}\label{eqn:laplacetransform_2ndTier_outer_2}
    & \mathcal{L}_{I_1 \mid X_2 = x, o \in A_{outer}}\big(\frac{T x^{\alpha}}{P_2} \big) \nonumber \\
    & =  \begin{cases}
      \exp \Big(- 2 \pi \lambda'_1 \int_D^{\infty} \big(1 - \frac{1}{1+x^{\alpha} (\frac{P_1}{P_2}) T y^{-\alpha}  } \big) y \mathrm{d}y \Big) \\
      \ \ \ \ \ \ \ \ \ \ \ \ \ \ \ \ \ \ \ \ \ \ \ \ \ \ \ \ \ \ \ \ \ \ \ \ \ \ \ \ \text{for $x \leqslant \big(\frac{P_2}{P_1}\big)^{1/\alpha} D $}\\
      \exp \Big(- 2 \pi \lambda'_1 \int_{x (\frac{P_1}{P_2})^{\frac{1}{\alpha}}}^{\infty} \big(1 - \frac{1}{1+x^{\alpha} (\frac{P_1}{P_2}) T y^{-\alpha}  } \big) y \mathrm{d}y \Big)\\
      \ \ \ \ \ \ \ \ \ \ \ \ \ \ \ \ \ \ \ \ \ \ \ \ \ \ \ \ \ \ \ \ \ \ \ \ \ \ \ \ \text{for $x > \big(\frac{P_2}{P_1}\big)^{1/\alpha} D $}\\
    \end{cases} \nonumber \\
    & =  \begin{cases}
      \exp \Big(- \pi \lambda'_1 D^2 \rho(\frac{P_1 T x^{\alpha}}{P_2 D^{\alpha}} , \alpha) \Big)
      & \text{for $x \leqslant \big(\frac{P_2}{P_1}\big)^{1/\alpha} D $}\\
      \exp \Big(- \pi \lambda'_1 (\frac{P_1}{P_2})^{2/\alpha} \rho(T,\alpha) x^2 \Big)
      & \text{for $x > \big(\frac{P_2}{P_1}\big)^{1/\alpha} D $},\\
    \end{cases}\nonumber \\
  \end{align}
Subsequently, we can substitute (\ref{eqn:laplacetransform_2ndTier_outer}), (\ref{eqn:laplacetransform_2ndTier_outer_2}) and the PDF of $X_2$, conditioned on that the typical user is in the outer region and provided in (\ref{eqn:pdf_X2_cond_outerarea}), into (\ref{eqn:probcov_2ndTier_outer_derive}). Consequentially, we can obtain the expression of the coverage probability in (\ref{eqn:probcov_2ndTier_outer}).

By combining the results in (\ref{eqn:probcov_1stTier_outer}) and (\ref{eqn:probcov_2ndTier_outer}), we can reach the coverage performance for a randomly chosen outer region typical user, provided by (\ref{eqn:probcov_outer}). Till here, we complete the proof.

\subsection{Proof of Theorem~\ref{theorem:Thput_cov_innerarea}} \label{appendix:proof_prop_Thput_cov_innerarea}

The CCDF of the throughput achieved at the inner region typical user, $\mathbb{P}[\mathcal{R} > \rho \mid o \in A_{inner}]$, can be obtained by assuming that all inner region users are served by the macrocell tier, that is,
  \begin{align}\label{eqn:Thput_cov_innerarea_derive1}
    \mathbb{P}&[\mathcal{R} > \rho \mid o \in A_{inner}]  \nonumber \\
    & \approx \mathbb{P}[\mathcal{R} > \rho \mid \omega = 1, o \in A_{inner}] \nonumber \\
    & = \mathbb{P}\Big[\frac{W}{N_{1} + 1} \log_2(1+\mathrm{SINR}_1) > \rho \mid  o \in A_{inner} \Big] \nonumber \\
    & = \mathbb{P}\big[\mathrm{SINR}_1 > 2^{(N_1 + 1)\rho /W} -1 \mid o \in A_{inner} \big] \nonumber \\
    & \stackrel{(a)}{\approx} \mathbb{E}_{N_{1}}\big[ p_{c, A_{inner}} \big(2^{(N_{1} + 1)\rho /W} -1 \big) \big] \nonumber \\
    & = \sum_{n=0}^{\infty} \mathbb{P}[N_{1} = n] \cdot p_{c, A_{inner}} \big(2^{(n + 1)\rho /W} -1 \big),
  \end{align}
where $N_1$ is the number of in-cell macrocell MSs sharing the resource with the typical user, and step $(a)$ is approximated by assuming the total independence between the distribution of $\mathrm{SINR}_{1}$ and $N_{1}$. As indicated in Appendix~\ref{appendix:proof_lemma_active_density}, the area of the macrocells can be approximated by $\mathcal{C}_{1} \approx \mathcal{C}_0 \big({\lambda_1}/{\mathcal{Q}_{1}}\big)$, which helps us to reach the PMF of $N_1$ in (\ref{eqn:PMF_MS_num_tier1}).

\subsection{Proof of Theorem~\ref{theorem:Thput_cov_outerarea}} \label{appendix:proof_prop_Thput_cov_outerarea}
\ignore{This result $\mathcal{Q}_{2,outer} = \mathbb{P}[\kappa = 2 \mid o \in A_{outer}]$ can be obtained by $\mathcal{Q}_{2,outer} = 1 - {\mathbb{P}[\kappa = 1, o \in A_{outer}]}/{\mathbb{P}[o \in A_{outer}]}$, in which $\mathbb{P}[\kappa = 1, o \in A_{outer}]$ and $\mathbb{P}[o \in A_{outer}]$ are provided in (\ref{eqn:prob_1stTier_derive2}) and (\ref{eqn:Prob_A_outer}). Hence, $\mathcal{Q}_{1,outer}$ can be easily obtained because of $ \mathcal{Q}_{1,outer}= 1 - \mathcal{Q}_{2,outer}$.}

The CCDF of the throughput achieved at the outer region typical user, $\mathbb{P}[\mathcal{R} > \rho \mid o \in A_{outer}]$, can be provided by
  \begin{align}\label{eqn:Thput_cov_outerarea_derive1}
    & \mathbb{P}[\mathcal{R} > \rho \mid o \in A_{outer}]  \nonumber \\
    & = \sum_{i \in \{1,2\}} \mathbb{P}[\mathcal{R} > \rho \mid \omega = i, o \in A_{outer}] \cdot \mathcal{Q}_{i,outer} \nonumber \\
    & \stackrel{(a)}{\approx} \sum_{i \in \{1,2\}} \mathbb{E}_{N_{i}}\big[ p_{c, i, A_{outer}} \big(2^{(N_{i} + 1)\rho /W} -1 \big) \big] \cdot \mathcal{Q}_{i,outer} \nonumber \\
    & = \sum_{i \in \{1,2\}} \sum_{n=0}^{\infty} \mathbb{P}[N_{i} = n]  p_{c, i, A_{outer}} \big(2^{(n + 1)\rho /W} -1 \big) \mathcal{Q}_{i,outer},
  \end{align}
where $N_i$ is the number of in-cell MSs sharing the resource with the typical user served by the $i$-th tier, and step $(a)$ is approximated by assuming the total independence between the distribution of $\mathrm{SINR}_{i}$ and $N_{i}$. As indicated in Appendix~\ref{appendix:proof_lemma_active_density}, the area of the small cells can be approximated by $\mathcal{C}_{2} \approx \mathcal{C}_0 \big({\lambda_2}/{\mathcal{Q}_{2,outer}}\big)$, which helps us to reach the PMF of $N_2$ in (\ref{eqn:PMF_MS_num_tier2}).






\bibliographystyle{IEEEtran}
\bibliography{IEEEabrv,Bib_Database}

\begin{thebibliography}{10}
\providecommand{\url}[1]{#1}
\csname url@rmstyle\endcsname
\providecommand{\newblock}{\relax}
\providecommand{\bibinfo}[2]{#2}
\providecommand\BIBentrySTDinterwordspacing{\spaceskip=0pt\relax}
\providecommand\BIBentryALTinterwordstretchfactor{4}
\providecommand\BIBentryALTinterwordspacing{\spaceskip=\fontdimen2\font plus
\BIBentryALTinterwordstretchfactor\fontdimen3\font minus
  \fontdimen4\font\relax}
\providecommand\BIBforeignlanguage[2]{{%
\expandafter\ifx\csname l@#1\endcsname\relax
\typeout{** WARNING: IEEEtran.bst: No hyphenation pattern has been}%
\typeout{** loaded for the language `#1'. Using the pattern for}%
\typeout{** the default language instead.}%
\else
\language=\csname l@#1\endcsname
\fi
#2}}

\bibitem{WanZho13ICC}
H.~Wang, X.~Zhou, and M.~C. Reed, ``Analytical evaluation of coverage-oriented
  femtocell network deployment,'' in \emph{Proc. IEEE Int'l Conf. on Commun.
  (ICC'13)}, Budapest, Hungary, June 2013, pp. 5974--5979.

\bibitem{And13MCOM}
J.~G. Andrews, ``Seven ways that {HetNets} are a cellular paradigm shift,''
  \emph{{IEEE} Commun. Mag.}, vol.~51, no.~3, pp. 136--144, 2013.

\bibitem{AndCla12JSAC}
J.~G. Andrews, H.~Claussen, M.~Dohler, S.~Rangan, and M.~C. Reed, ``Femtocells:
  {Past}, present, and future,'' \emph{{IEEE} J. Select. Areas Commun.},
  vol.~30, no.~3, pp. 497--508, Apr. 2012.

\bibitem{ChaKou09TWC}
V.~Chandrasekhar, M.~Kountouris, and J.~G. Andrews, ``Coverage in multi-antenna
  two-tier networks,'' \emph{{IEEE} Trans. Wireless Commun.}, vol.~8, no.~10,
  pp. 5314--5327, Oct. 2009.

\bibitem{WeiGro10MCOM}
J.~Weitzen and T.~Grosch, ``Comparing coverage quality for femtocell and
  macrocell broadband data services,'' \emph{{IEEE} Commun. Mag.}, vol.~48,
  no.~1, pp. 40--44, Jan. 2010.

\bibitem{JoSan12TWC}
H.-S. Jo, Y.~J. Sang, P.~Xia, and J.~G. Andrews, ``Heterogeneous cellular
  networks with flexible cell association: {A} comprehensive downlink {SINR}
  analysis,'' \emph{{IEEE} Trans. Wireless Commun.}, vol.~11, no.~10, pp.
  3484--3495, 2012.

\bibitem{DhiGan12JSAC}
H.~S. Dhillon, R.~K. Ganti, F.~Baccelli, and J.~G. Andrews, ``Modeling and
  analysis of {K}-tier downlink heterogeneous cellular networks,'' \emph{{IEEE}
  J. Select. Areas Commun.}, vol.~30, no.~3, pp. 550--560, Apr. 2012.

\newpage
\bibitem{WanRee12AusCTW}
H.~Wang and M.~C. Reed, ``Tractable model for heterogeneous cellular networks
  with directional antennas,'' in \emph{Proc. 2012 Australian Commun. Theory
  Workshop (AusCTW'12)}, Wellington, New Zealand, Jan./Feb. 2012, pp. 61--65.

\bibitem{StoKen95Book}
D.~Stoyan, W.~S. Kendall, and J.~Mecke, \emph{Stochastic Geometry and its
  Applications}, 2nd~ed.\hskip 1em plus 0.5em minus 0.4em\relax New York, NY:
  John Wiley \& Sons Ltd., 1995.

\bibitem{BacBla09Book}
F.~Baccelli and B.~B{\l}aszczyszyn, \emph{{Stochastic Geometry and Wireless
  Networks, Volume I: Theory}}.\hskip 1em plus 0.5em minus 0.4em\relax Hanover,
  MA: Now Publishers Inc., 2009.

\bibitem{Bro00JSAC}
T.~X. Brown, ``Cellular performance bounds via shotgun cellular systems,''
  \emph{{IEEE} J. Select. Areas Commun.}, vol.~18, no.~11, pp. 2443--2455, Nov.
  2000.

\bibitem{YanPet03TSP}
X.~Yang and A.~P. Petropulu, ``Co-channel interference modeling and analysis in
  a {Poisson} field of interferers in wireless communications,'' \emph{{IEEE}
  Trans. Signal Processing}, vol.~51, no.~1, pp. 64--76, Jan. 2003.

\bibitem{Hae08TIT}
M.~Haenggi, ``A geometric interpretation of fading in wireless networks:
  {Theory} and applications,'' \emph{{IEEE} Trans. Inform. Theory}, vol.~54,
  no.~12, pp. 5500--5510, Dec. 2008.

\bibitem{AndBac11JCOM}
J.~G. Andrews, F.~Baccelli, and R.~K. Ganti, ``A tractable approach to coverage
  and rate in cellular networks,'' \emph{{IEEE} Trans. Commun.}, vol.~59,
  no.~11, pp. 3122--3134, Nov. 2011.

\bibitem{NovGan11TWC}
T.~D. Novlan, R.~K. Ganti, A.~Ghosh, and J.~G. Andrews, ``Analytical evaluation
  of fractional frequency reuse for {OFDMA} cellular networks,'' \emph{{IEEE}
  Trans. Wireless Commun.}, vol.~10, no.~12, pp. 4294--4305, Dec. 2011.

\bibitem{WanQue12JSAC}
C.~C. Wang, T.~Q.~S. Quek, and M.~Kountouris, ``Throughput optimization,
  spectrum allocation, and access control in two-tier femtocell networks,''
  \emph{{IEEE} J. Select. Areas Commun.}, vol.~30, no.~3, pp. 561--574, Apr.
  2012.

\bibitem{CheNgu12VTC}
C.~S. Chen, V.~M. Nguyen, and L.~Thomas, ``On small cell network deployment: A
  comparative study of random and grid topologies,'' in \emph{Proc. {IEEE} 76th
  Vehic. Tech. Conf. (VTC'12-Fall)}, Qu\'{e}bec City, Canada, Sept. 2012, pp.
  1--5.

\bibitem{YuKim11arXiv}
\BIBentryALTinterwordspacing
S.~M. Yu and S.-L. Kim. Downlink capacity and base station density in cellular
  networks. [Online]. Available: \url{http://arxiv.org/abs/1109.2992}
\BIBentrySTDinterwordspacing

\bibitem{SinDhi12arXiv}
\BIBentryALTinterwordspacing
S.~Singh, H.~S. Dhillon, and J.~G. Andrews. Offloading in heterogeneous
  networks: {Modeling}, analysis and design insights. [Online]. Available:
  \url{http://arxiv.org/abs/1208.1977}
\BIBentrySTDinterwordspacing

\bibitem{Hae13arXiv}
\BIBentryALTinterwordspacing
M.~Haenggi. A versatile dependent model for heterogeneous cellular networks.
  [Online]. Available: \url{http://arxiv.org/abs/1305.0947}
\BIBentrySTDinterwordspacing

\bibitem{BlaKar13Infocom}
B.~B{\l}aszczyszyn, M.~K. Karray, and H.~P. Keeler, ``Using {Poisson} processes
  to model lattice cellular networks,'' in \emph{Proc. 32nd {IEEE} Int'l Conf.
  on Computer Communications (INFOCOM'13)}, Turin, Italy, Apr. 2013, pp.
  773--781.

\bibitem{NiuLee13TWC}
J.~Niu, D.~Lee, X.~Ren, G.~Y. Li, and T.~Su, ``Scheduling exploiting frequency
  and multi-user diversity in {LTE} downlink systems,'' \emph{{IEEE} Trans.
  Wireless Commun.}, vol.~12, no.~4, pp. 1843--1849, 2013.

\bibitem{DahPar08HSPALTEbook}
E.~Dahlman, S.~Parkvall, J.~Skold, and P.~Beming, \emph{{3G} Evolution: {HSPA}
  and {LTE} for Mobile Broadband}, 2nd~ed.\hskip 1em plus 0.5em minus
  0.4em\relax Burlington, MA: Academic Press, 2008.

\bibitem{OkaBoo92Book}
A.~Okabe, B.~Boots, and K.~Sugihara, \emph{Spatial Tessellations: Concepts and
  Applications of {Voronoi} Diagrams}.\hskip 1em plus 0.5em minus 0.4em\relax
  New York, NY: John Wiley \& Sons Ltd., 1992.

\bibitem{HinMil80Math}
A.~L. Hinde and R.~E. Miles, ``{Monte Carlo} estimates of the distributions of
  the random polygons of the {Voronoi} tessellation with respect to a {Poisson}
  process,'' \emph{Journal of Statistical Computation and Simulation}, vol.~10,
  no. 3-4, pp. 205--223, 1980.

\bibitem{WeaKer86Math}
D.~Weaire, J.~P. Kermode, and J.~Wejchert, ``On the distribution of cell areas
  in a {Voronoi} network,'' \emph{Philosophical Magazine Part B}, vol.~53,
  no.~5, pp. L101--L105, 1986.

\bibitem{LeeHae12TWC}
C.-H. Lee and M.~Haenggi, ``Interference and outage in {Poisson} cognitive
  networks,'' \emph{{IEEE} Trans. Wireless Commun.}, vol.~11, no.~4, pp.
  1392--1401, 2012.

\bibitem{EftMac06TNSPS}
M.~Efthymiou, A.~Mackay, A.~Dow, and M.~Flanagan, ``Spatial optimisation: {How}
  subscribers can help you optimize your {CDMA} network,'' in \emph{Proc. 12th
  Int'l Telecom. Network Strategy and Planning Symp. (Networks'06)}, New Delhi,
  India, Nov. 2006, pp. 1--11.

\end{thebibliography}

\end{document}